# The Effect of Junction Gutters for the Upscaling of Droplet Generation in a Microfluidic T-Junction

## H. Viswanathan*

*Department of Engineering and Mathematics, Sheffield Hallam University, Howard Street, Sheffield, England, S1 1WB, United Kingdom*

## ABSTRACT

The influence of drop formation due to micro rib-like structures, viz., the Junction Gutters (JGs) within a standard microfluidic T-junction, is numerically investigated. Hydrodynamic conditions that lead to various flow regimes are identified characterized by the Capillary number ($Ca$) and velocity ratios of the dispersed and continuous phases ($q$) within a standard T-junction. Subsequently, under such conditions, a range of gutter configurations is introduced in the standard channel. The results predict that the introduction of JGs can favourably alter the formation frequency and morphology of drops and, consequently, promote upscaling significantly for the hydrodynamic conditions associated with low $Ca$. Detailed flow maps are presented that reveal a plethora of transitions during the formation of droplets with higher $Ca$ and $q$ that would otherwise signify a dripping or a jetting regime in a standard junction. However, specific gutter configurations are identified where JGs are unfavourable for generating monodisperse droplets.

**\*Corresponding author information:** Email: h.viswanathan@shu.ac.uk ; Phone: +0441142256244

# Abbreviations

| | |
|---|---|
| $a, b$ | Length and depth of junction gutter |
| $a^*, b^*$ | Dimensionless length and depth of junction gutter |
| $W_c$ | Width of continuous phase inlet (and the main channel) |
| $W_d$ | Width of dispersed phase inlet |
| $L_M$ | Length of the main channel |
| $L_{ec}, L_{ed}$ | Entrance Lengths of continuous and dispersed phase channels |
| $Ca$ | Capillary number |
| $q$ | velocity ratio between dispersed and continuous phases. |
| $P_j$ | Junction pressure |
| $Re$ | Reynolds number in the Channel |
| $C_{b1}, C_{b2}$ | Boundaries of the main channel. |
| $L_D{}^*$ | Dimensionless size of droplet |
| $f$ | formation frequency of droplet |
| $t^*$ | time |
| $x, y$ | Coordinate Systems |
| $\beta, \gamma$ | Fitting parameters |
| $\theta$ | Static contact angle |
| $\eta$ | Viscosity ratio |
| $\rho_c, \rho_d$ | Density of continuous and dispersed phase |
| $\mu_c, \mu_d$ | Viscosity of continuous and dispersed phase |
| $\sigma$ | Surface tension coefficient |
| $\alpha$ | Volume fraction |
| $\Psi$ | Channel aspect ratio between the width of dispersed and continuous phases |
| $\lambda$ | Droplet spacing |
| JG | Junction Gutter |
| VOF | Volume of Fluid |

| CLSVOF | Coupled Level Set -and Volume of Fluid |
| ITA | Iterative Algorithm |
| NITA | Non-Iterative Algorithm |

## 1. Introduction

Droplet-based microfluidics is continuously evolving with applications associated with several aspects of science and engineering due to the reliable manipulation of drops (Zhao and Middelberg, 2011, Stone et al., 2004, Link et al., 2004). These include chemical and medical applications wherein highly uniform droplet formation is a constant requirement, such as compartmentalization in biological assays (Scheler et al., 2018), medical diagnostics (Rivet et al., 2011), cell-encapsulation (Köster et al., 2008), DNA-sequencing (Lan et al., 2017) and drug release for which microfluidic Lab-on-Chip devices are employed (Cui and Wang, 2019). More recently, microfluidic devices have begun to evolve as excellent platforms for detecting RNA viruses (Basiri et al., 2020, Dolan et al., 2018). However, despite such substantial breakthroughs achieved in the society using microfluidic technologies, there are significant underlying challenges wherein *a)* droplet production rates using microfluidic devices owing to exorbitant handling costs and requirements on *b)* highest precision, control, and stringent quality standards during fabrication of microfluidic devices, bearing in mind their life-saving applications. To complicate matters further, it becomes essential to ensure the desired chemical and biological transformations during droplet fragmentation are intrinsically safe and environmentally friendly.

Typically, in two-phase, liquid-liquid microfluidic systems within the scope of applications mentioned above, to precisely control, manipulate and enhance the droplet throughput, conditions such as the flow behaviour, droplet size, conditions of wettability and the geometry of the microfluidic device become critical (Sattari et al., 2020). Therefore, various microfluidic channels exist, such as but are not restricted to i) cross-flow, ii) co-flow, and c) flow-focusing and several geometric variations within them are possible. The T-junction is a form of a cross-flow configuration where the dispersed and the continuous phase fluids are fed orthogonally to generate droplets. Since its inception (Thorsen et al., 2001), the T-junction has significantly gained popularity owing to its simplicity and ability to produce monodisperse droplets with a coefficient of variation of <2% (Xu et al., 2006). Furthermore, considering its capability to robustly upscale through minimal modifications through methods such as integrating several parallel-T junctions (Nisisako and Torri, 2008), split the primary and secondary droplets (Sun et al., 2018). More recently, their potential to be configured within an Interactive Learning Control (ILC) (Huang et al., 2020) makes T-junctions favourable for mass production of droplets with high break up rates (Zhu and Wang, 2017).

Nevertheless, the complexities involved with the dynamics of the fluids and their interactions with the T-junction configuration and a constant need for device miniaturization, scaling and upscaling continue to pose challenges associated with achieving the control and breakup of droplets (Chiu et al., 2017). Therefore, to overcome such challenges, several passive and active methods have been proposed where the former does not require external actuation. In contrast, the latter typically makes use of additional energy through electrical (Singh et al., 2020), thermal (Murshed et al., 2008), magnetic (Tan and Nguyen, 2011), and mechanical actuation (Churski et al., 2010) with which the droplets are generated within the framework of a T-junction configuration. Although most active methods yield an excellent coefficient of variation and present a range of possibilities for drop generation, the challenges for parallelization, an additional level of complexity in handing the external input, and cost-based constraints may persist depending on the nature of the external input employed (Chong et al., 2016). In passive methods, the hydrodynamic conditions such as capillary number (Ca), the flow rate or velocity ratios of the dispersed and continuous phases, and at times, trivial geometric changes to the standard cross-flow or a co-flow configuration can be harnessed to generate droplets that emanate from flow regimes in a T-junction such as squeezing, dripping, and jetting (De Menech et al., 2008, Li et al. 2012).

Several research works have successfully modified the standard T-junction configuration both experimentally and numerically to enhance the process of drop formation through passive methods. Abate and Weitz (2011) experimentally proposed a modified T-junction that consisted of three junctions; a jetting junction, a bubbling junction where air bubbles were formed and a triggering junction in which the air bubbles deformed the jet to form droplets due to Rayleigh-Plateau instability. Shui et al. (2009) developed a head-on T-junction configuration in which two identical inlet channels, a constriction channel a wide outlet channel to examine the drop formation at different flow regimes. Their results suggested that the head-on devices have wider applicability to generate a broad range of droplet sizes in regimes driven by capillary instability, squeezing and shearing. Various numerical studies on the modified T-junctions have evolved in the recent past that investigated the head-on T junction in the form of an orthogonal double junction (Ngo et al., 2015; Han and Chen, 2019, Raja et al., 2021) to generate alternating droplets and investigated the resulting drop sizes due to channel width and with different junction injection angles in standard T-junctions (Jamalabadi et al., 2017) and for double junctions ranging between 30-90 degrees (Ngo et al., 2016) that suggested the drop formation in an alternating pattern increases with injection angles. Consequently, the studies described above suggest that topological changes to the standard T-junction can be utilized to significantly promote droplet/ bubble breakup (Arias, 2020), as detailed in the review of Cerderia et al. (2020).

Recently, a step like modification and a capillary device in the standard T-junction has shown to have substantial potential to produce monodisperse droplets under jetting regimes wherein polydisperse droplets are most often realized due to the uncontrollable Rayleigh instability (Li et al., 2015). However, with the step-like modification to the standard T-junction in place, the experiments of Li et al. (2015) demonstrated that monodisperse microdrops are accomplished owing to the narrowing jetting flow that can be realized. It was further demonstrated that monodisperse drops as small as 15 µm could be achieved under stable jetting conditions agreeing with the theoretical scaling. (Li et al., 2016). The most recent work of Zhang et al. (2022) that implemented the deep learning technique to enhance the ease of measuring the microdrops formed with the step arrangement in the T-junction further endorses the efficacy of such a step-like arrangement for forming monodisperse drops passively. The use of rib-like structures within the T-junction channel similar to the step arrangement mentioned above was numerically investigated comprehensively by Li et al., 2019. Four different rib structures viz., two triangular structures that are of the same width and height but with different orientations in reference to upstream and downstream of the channel, a streamlined structure formed by a semicircle complemented by a quarter of a circle, and a rectangular structure were embedded in the junction. The superiority of the rectangular rib was adequately demonstrated in their work which suggested that in the flow regime phase map, the jetting regime is curtailed by the rib, which resulted in favourably expanding the phase space of the dripping regime. New scaling laws for squeezing and dripping regimes were derived that indicated that the droplet diameter from the T-junction with the rectangular rib decreases linearly with the micro rib intrusion into the channel. The work of Li et al. (2019) opens further avenues for exploration of the rib-like structure closer to the junction, which acts as a droplet gutter by effectively promoting passive droplet generation in T-Junctions.

The present numerical work developed in this paper derives motivations from the works of Li et al. (2019), Li et al. (2015), and Li et al. (2016). The current work furthers the investigation by exploring *i)* the behaviour of a wide range of rectangular/square junction gutters when embedded onto the standard T-junction subjected to various flow regimes, *ii)* transitions during drop formation that occur under the same hydrodynamic conditions within various flow regimes by purely varying the gutter topologies, and *iii)* the flow regime maps of the gutter phase-space topologies that promote and deteriorate droplet upscaling are identified.

## 2. System Details and Numerical Method

### 2.1 System Description

The schematic of the microfluidic T-junction together with the Junction Gutter (JG) of length ($a$) and depth ($b$) is described in Fig.1. A two-dimensional modelling approach is chosen for this study wherein the continuous phase inlet flows through the main channel and interacts with the dispersed phase injected through the side channel. The width of the continuous phase inlet and the main channel ($W_c$), the width of the dispersed phase inlet ($W_D$), the fluid properties of the continuous and dispersed phases are maintained as identical to that presented by Glawdel et al. (2012) since the current numerical work is validated against their experiments as demonstrated in Section 3.

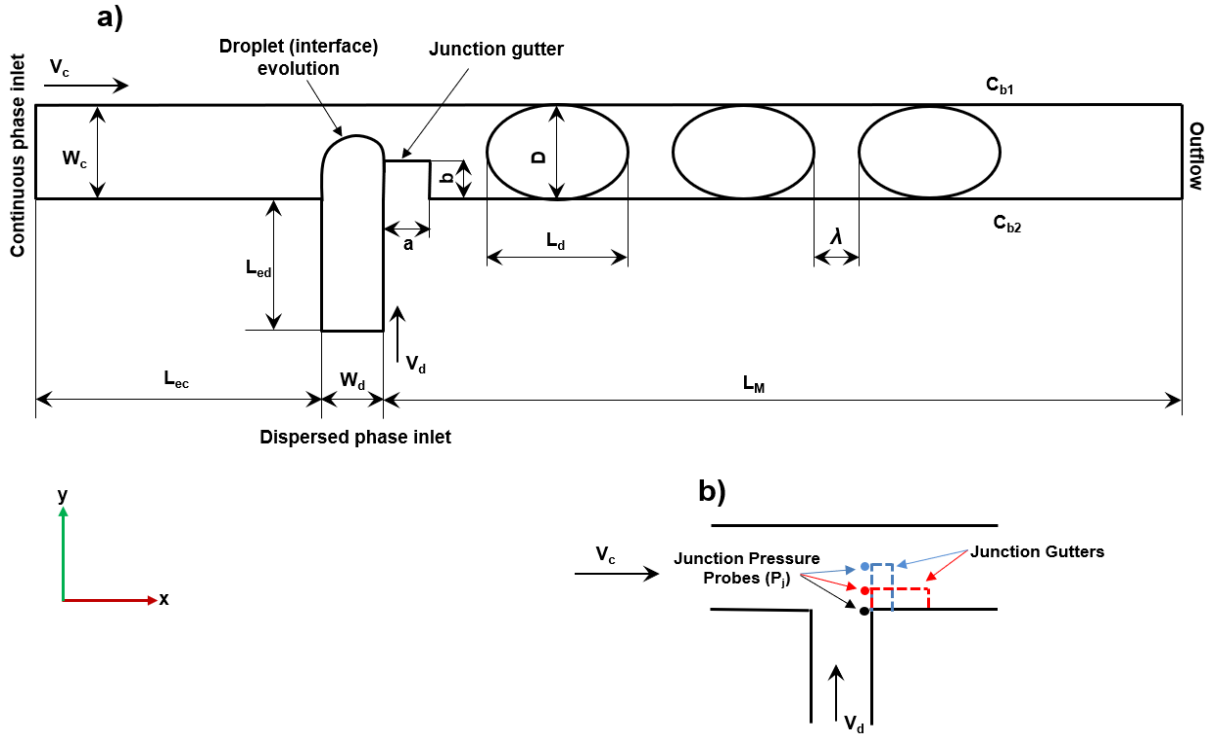

**Fig. 1**. Schematic representations of **a)** the microfluidic T-junction, gutter position and boundary conditions, **b)** location of the junction pressure probes ($P_j$) w.r.t the standard T-junction and different gutter arrangements.

To ensure a fully developed laminar flow, the entrance lengths of the main channel ($L_{ec}$) and the side channels ($L_{ed}$) are chosen in accordance with Eq.1. (Nekouei and Vanapalli, 2017) that is used previously by Li et al. (2019):

$$L_{ec} = W_c \left( \frac{0.6}{1+0.035Re} + 0.056Re \right), \qquad (1)$$

Where $W_c$ corresponds to the hydraulic diameter, which is essentially the width of the continuous phase channel, and $Re$ corresponds the Reynolds number. The Eq.1. stated herein is for the continuous phase entrance length; however, it takes the same form with the corresponding parameters to obtain the dispersed phase entrance. The chosen entrance lengths correspond to the largest $Re$ that is investigated in this study for both the continuous and dispersed phase channels, respectively, which is in line with the work of Li et al. 2019. The dimensions of the microchannel and the properties of the fluids used in the current work is presented in Table 1.

**Table 1.** Description of Microchannel, gutter dimensions and liquid properties used in the simulations.

| Sc. no | Dimensions and Properties | Values |
|--------|---------------------------|--------|
| i) | Width of continuous phase inlet (and main channel) ($W_c$) | 100 μm |
| ii) | Width of dispersed phase inlet ($W_d$) | 90 μm |
| iii) | Length of the main channel ($L_M$) | 4600 μm |
| iv) | Entrance length of the continuous phase channel ($L_{ec}$) | 800 μm |
| v) | Entrance length of the dispersed phase channel ($L_{ed}$) | 170 μm |
| vi) | Gutter Length ($a$) | 5- 100 μm |
| vii) | Gutter Depth ($b$) | 10- 80 μm |
| viii) | Density of continuous phase ($\rho_c$) | 930 kg/m$^3$ |
| ix) | Density of dispersed phase ($\rho_d$) | 1024 kg/m$^3$ |
| x) | Viscosity of continuous phase ($\mu_c$) | 10.2 mPa s |
| xi) | Viscosity of dispersed phase ($\mu_d$) | 1.21 mPa s |
| xii) | Surface tension coefficient ($\sigma$) | 0.0372 N/m |

## 2.2 Numerical Description and Boundary Conditions

A range of numerical approaches such as the level set method (LSM) (Bashir et al. 2011, Wong et al. 2017), the volume of fluid (VOF) (Kashid et al. 2010, Ngo et al., 2015; Mastiani et al. 2017; Zhang et al. 2015), coupled- level set- and volume of fluid (CLSVOF) (Chakraborty et al., 2019) lattice Boltzmann (LBM) (Liu and Zhang, 2009; Chen and Deng, 2017) have been successfully employed in describing various regimes of drop formation in a microfluidic T-junction within the 2D framework. An overview of the numerical techniques, together with their advantages and disadvantages, are comprehensively described in the work of van Sit Annaland et al. (2005). The VOF is a free surface reconstruction method that offers a simpler but robust treatment of the topological changes of the interface (Viswanathan, 2019, Viswanathan, 2020) and can be applied for effectively describing droplet breakup and coalescence. However, the CLSVOF is a hybrid approach that couples the level set function to the VOF to estimate the curvature of the interface more adequately.

In the present work, both the VOF and the CLSVOF are assessed within a standard T-junction (presented in **Appendix A)**, and their suitability is examined for their applicability into T-junctions embedded with JGs. The equations that govern the flow in the system are described as follows:

The continuity equation is given as

$$\frac{\partial \rho}{\partial t} + \boldsymbol{\nabla} \cdot (\rho \vec{v}) = 0 \qquad (2)$$

and the momentum equation is described by

$$\frac{\partial}{\partial t}(\rho \vec{v}) + \boldsymbol{\nabla} \cdot (\rho \vec{v} \vec{v}) = -\nabla p + \boldsymbol{\nabla} \cdot [\mu(\boldsymbol{\nabla}\vec{v} + \boldsymbol{\nabla}\vec{v}^T)] + \vec{F} \qquad (3)$$

where $\vec{v}$ is velocity, $\rho$ is the density, $p$ is the pressure, $\mu$ is the viscosity, and $t$ is the time. $\vec{F}$ is the continuum surface tension force on the interface (Brackbill et al. 1992) of the volume fraction field $\alpha$, given by

$$\vec{F} = \sigma k \nabla \alpha \qquad (4)$$

where $k$ is the local curvature on the interface and is computed as

$$k = -\nabla \cdot \left( \frac{\nabla \alpha}{|\nabla \alpha|} \right) \qquad (5)$$

In terms of describing the interface between two immiscible fluids, namely, the continuous and dispersed phases, and providing the volume fraction conservation throughout the domain, a two-phase flow description of the Volume of Fluid method (VOF) is incorporated. The VOF equation is given by:

$$\frac{\partial \alpha}{\partial t} + \nabla \cdot (\vec{v}\alpha) = 0 \qquad (6)$$

The volume fraction gives the portion of the cell which is filled with either phase, where

$$
\begin{array}{ll}
\alpha = 0 & \text{the cell is filled with the continuous phase fluid} \\
0 < \alpha < 1 & \text{the interface exists in the cell} \\
\alpha = 1 & \text{the cell is filled with the dispersed phase fluid}
\end{array} \qquad (7)
$$

The density $\rho$ and viscosity $\mu$ can be expressed as linear contributions from the continuous and dispersed phases indicated by the subscripts $d$ and $c$ as follows:

$$\rho = \rho_d \alpha + \rho_c (1 - \alpha) \qquad (8)$$
$$\mu = \mu_d \alpha + \mu_c (1 - \alpha) \qquad (9)$$

For the CLSVOF method, firstly the level set function is defined $\emptyset$ is defined by

$$\frac{\partial \emptyset}{\partial t} + \nabla \cdot (\vec{v}\emptyset) = 0 \qquad (10)$$

The liquid phase properties that are interpolated across the interface are given by

$$\rho = \rho_d H(\emptyset) + \rho_c (1 - H(\emptyset)) \qquad (11)$$
$$\mu = \mu_d H(\emptyset) + \mu_c (1 - H(\emptyset)) \qquad (12)$$

where the Heaviside function $H(\emptyset)$ can be written as follows (Sussman et al. 1994)

$$H(\emptyset) = \begin{cases} 0 & if \ \emptyset < -\varepsilon \\ \frac{1}{2} + \frac{\emptyset}{2\varepsilon} + \frac{1}{2\pi}\sin\frac{\pi\emptyset}{\varepsilon} & if \ |\emptyset| \leq \varepsilon \\ 1 & if \ \emptyset > \varepsilon \end{cases} \tag{13}$$

where $2\varepsilon$ is the finite interface thickness over which the fluid properties are smoothed. The value of $\varepsilon$ is typically chosen between one to four times the length of the smallest computational cell so that numerical instability owing to parasitic currents are prevented.

The surface tension force in the case of CLSVOF is given by,

$$\vec{F} = \sigma \alpha k \nabla H(\emptyset) \tag{14}$$

where the interface curvature is determined by

$$k = -\nabla \cdot \left( \frac{\nabla \emptyset}{|\nabla \emptyset|} \right) \tag{15}$$

In both cases, the wall adhesion is taken into account through a contact angle $\theta$ at the channel wall given by

$$\hat{n} = \hat{n}_w \cos\theta + \hat{t}_w \sin\theta \tag{16}$$

where $\hat{n}$ is the surface normal and $\hat{n}_w$, $\hat{t}_w$ are, unit vectors normal and tangential to the wall, respectively. In the present simulations, it is assumed that the continuous phase perfectly wets the wall of the channel walls. In all the simulations in this study, the value of $\theta$ is fixed to be at $140^0$. As shown in Fig.1, a uniform velocity is provided at the continuous and dispersed phase inlets, and at the channel boundaries $C_{b1}$ and $C_{b2}$, a no-slip condition is employed. A zero-pressure condition is prescribed at the outlet boundary. The dimensionless parameters that characterize this system are the Capillary number $Ca = \frac{\mu_c U_c}{\sigma}$, the channel Reynolds number $Re = \frac{\rho U_c W_c}{\mu_c}$, the viscosity ratio $\eta = \frac{\mu_d}{\mu_c}$, channel aspect ratio $\Psi = \frac{W_d}{W_c}$, velocity ratio $q = \frac{U_d}{U_c}$ where $U_d$ and $U_c$ correspond to the maximum velocity of the dispersed and continuous phases. Considering that the maximum Reynolds number $Re \ll 1$, and $\Psi$, $\eta$ are fixed as given in Table.1, therefore, the two governing parameters are $Ca$ and $q$ for a standard microchannel T-junction. However, the presence of JGs in this system introduces two additional dimensionless parameters viz., the dimensionless gutter length $a^* = \frac{a}{W_c}$ and depth $b^* = \frac{b}{W_c}$ respectively.

### 2.3 Numerical Solution Procedure

The governing equations described in Section 2.2 are solved using the commercially available finite-volume based commercial software, ANSYS Fluent, Version 2020 R2. For the VOF method, the pressure-velocity scheme used was PISO that splits the solution into predictor and corrector steps together with the Non-Iterative Algorithm (NITA). However, in the case of the CLSVOF method, the PISO scheme was used in conjunction with the Iterative Algorithm (ITA) that required at least 45 inner iterations to ensure that all residuals of the CLSVOF method meet the sufficient convergence tolerance as demonstrated by the VOF, i.e., $<10^{-6}$. In both cases, the momentum equation was discretized using the QUICK scheme, and the gradients of the scalars were computed by using the Least-Squares cell-based method. The Least-Squares cell-based was chosen since it is directly comparable to node-based gradient methods, is much more accurate than cell-based methods, and is computationally less intensive. The "PRESTO!" (PREssure STaggering Option) scheme, although computationally more expensive, was used to interpolate the pressure term as it directly calculates the pressures at cell faces and avoids interpolation errors. The interface was determined by the Geo-Reconstruct Algorithm (Youngs, 1982) that uses a piecewise-linear approach to determine the interface between fluids. To calculate the interfacial forces, the Continnum Surface Force (CSF) model was used (Brackbill et al., 1992). In the case of CLSVOF, the level set function described by Eq.10 was discretized using the QUICK scheme. In both cases, the time-step is chosen to ensure that the Courant number is lesser than 0.25. For more information on the VOF solution process, the readers are directed to references (Viswanathan, 2019, Viswanathan, 2020).

## 3. Grid Verification, Choice of Methods and Validation

The selection of a grid and the choice of methods was a multi-fold process. Firstly, the grid verification test is conducted by evaluating meshes of sizes 5 μm (coarse), 4 μm (medium), and 3 μm (fine), as shown in Fig. 2.

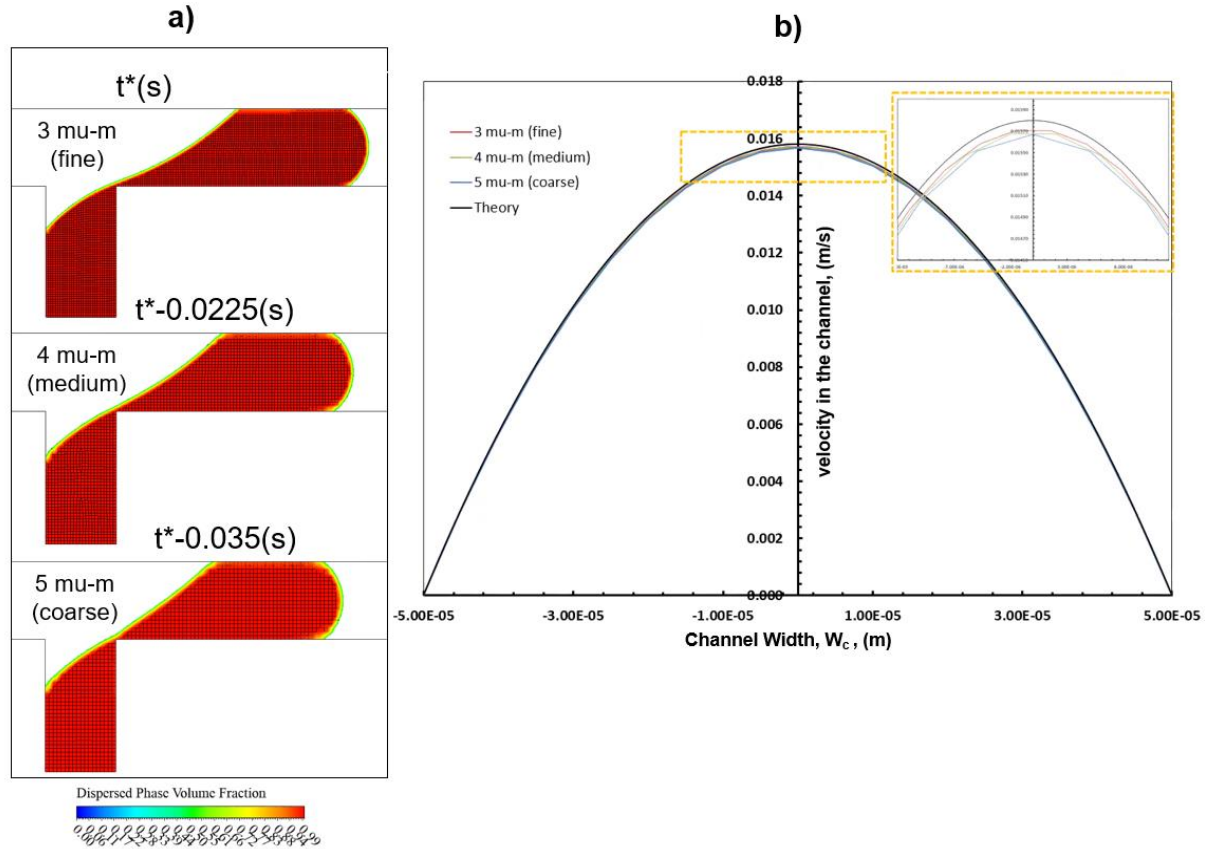

**Fig. 2**. Grid verification study for $Ca$ = 0.0043, $q$ = 0.6487: **a)** spatial and temporal evolutions of the interface for fine, medium, and coarse grids, **b)** variation of velocity along the channel at y= 0.625 $L_{ec}$ (See **Table. 1**. for details of $L_{ec}$) for different grids evaluated against the theoretical description of the fully developed flow.

The spatially evolved droplet profiles superimposed with the grid details, just before pinch-off, are shown for the three mesh cases in Fig. 2**a**). The curvature associated with the evolution of the neck for the 3 μm (fine) is more prominent compared to the other cases. The value associated with $t^*$ denotes the time just before the droplet is formed with 3 μm (fine) case and it is observed that time for formation of the droplet relatively increases with the increased mesh size. Further, the theoretical velocity profile $u_c(y)$ for a fully developed laminar flow based on entrance length estimate from Eq. 1 is given by.

$$u_c(y) = V_c \left(1 - \left(\frac{y}{W_c}\right)^2\right) \tag{17}$$

A comparison of the numerically predicted velocity profiles against the theory (Eq. 17) for the three mesh cases at the same location, is shown in Fig. 2**b**). The differences between the fine and medium grids appear to be much lesser compared to the coarse grid, which is consistent with the observation in Fig.2**a**). Although a good agreement is seen between the theory and numerical predictions for all the grids, the fine grid appears to show the least difference with the overall theoretical profile, and therefore, the 3 µm (fine) grid was employed for all the cases examined in this study. In the present work, a comparison between two interface capturing methods viz., the VOF and CLSVOF is presented in **Appendix A** to justify the choice of methods. As shown in Table 2, despite a close numerical prediction, the CLSVOF method, together with the iterative time advancement (ITA) scheme, requires ~3.6 times higher time to compute than compared to its VOF counterpart using the non- iterative time advancement (NITA) scheme. Therefore, the VOF method, together with the 3 µm (fine) grid alongside the numerical procedure described in Section 2.3, was employed throughout the rest of the analysis.

**Table. 2.** Grid verification study and a comparison between VOF and CLSVOF interface capturing methods against the experiment (Glawdel et al. 2012).

| Method | Grid sizes | Time advancement algorithm | Total simulation time normalized w.r.t VOF (fine grid) | Predicted drop formation frequency (f) Hz | $L_D^* = \dfrac{L_D}{W_c}$ |
|---|---|---|---|---|---|
| VOF | 5 µm (coarse) | NITA | 0.6747 | 30.7692 | 2.3760 |
| VOF | 4 µm (medium) | NITA | 0.8343 | 30.0751 | 2.4080 |
| VOF | 3 µm (fine) | NITA | 1.000 | 29.4117 | 2.4763 |
| CLSVOF | 3 µm (fine) | ITA | 3.6295 | 29.4621 | 2.4758 |
| Experiment (Glawdel et al. (2012)) | - | - | - | 29.7 | - |

The prediction from the VOF model is validated by comparing against the experimental data of Glawdel et al. (2012). Both the spatial and temporal periods of the drop formation and pinch-off such as the i) lag stage where the interface recedes to a small distance back into the dispersed phase inlet before it enters into the main channel, ii) filling stage wherein the interface penetrates and fills the main channel by proceeding towards the channel boundary $C_{b1}$, iii) transitioning into necking and iv) detachment are directly compared against the experimental data as shown in Fig.3**a)**. The results show that the numerical predictions agree well with the experiments at various stages of drop formation and the overall droplet dimensions; an additional validation reinforces this with a different dispersed phase inlet dimension and hydrodynamic conditions presented in **Appendix B**. Nevertheless, the necking behaviour observed in the experimental data appears more pronounced than the numerical results suggesting that the stage between necking and pinch-off is quicker with the model.

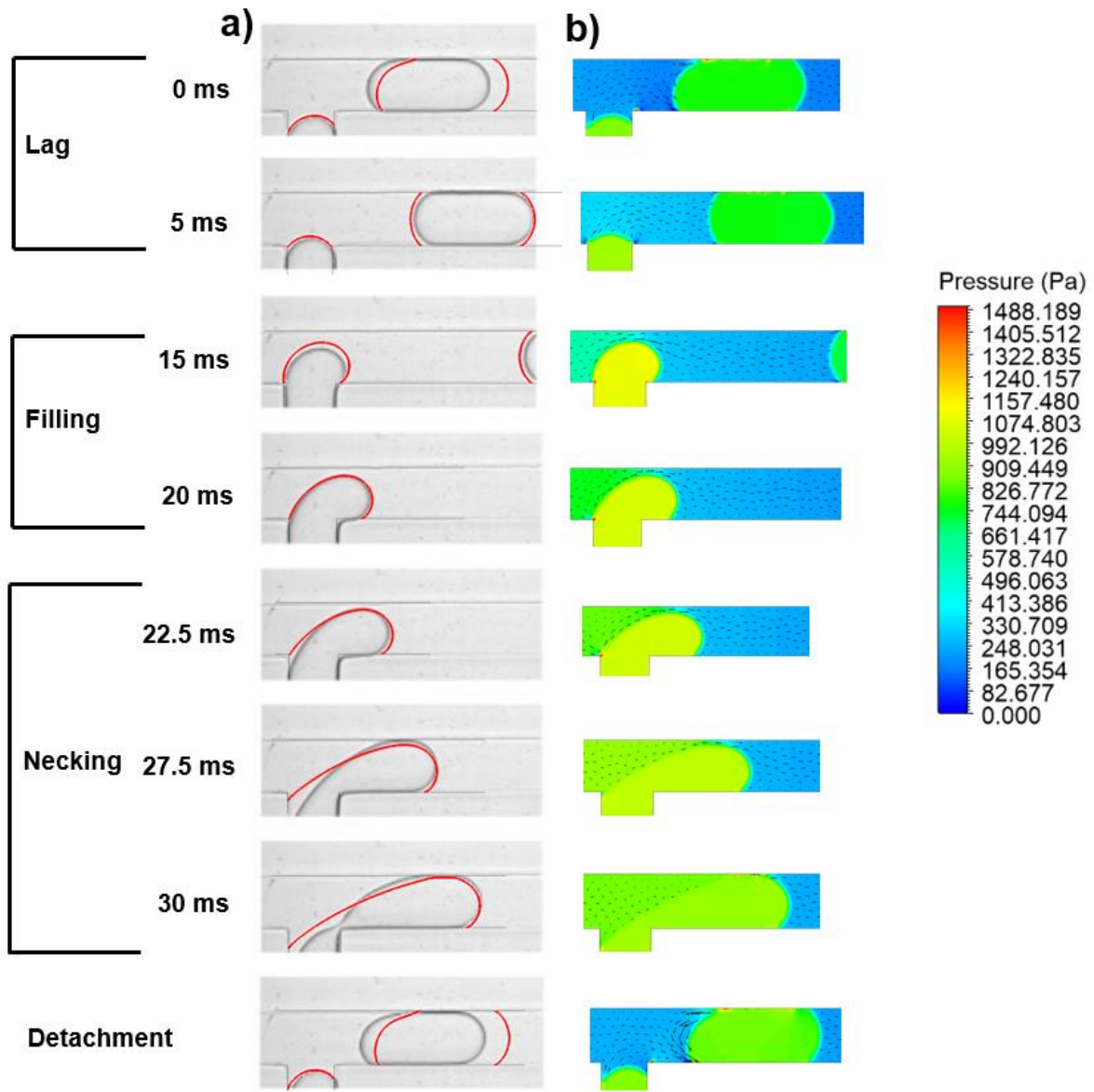

**Fig. 3**. **a)** Comparison between the experimental result of Glawdel et al. (2012) (Reproduced with permission from, Copyright 2012 APS) against the current numerical (VOF) predictions (shown by the solid red line) during different stages of drop formation for $Ca$ = 0.0043, $q$ = 0.6487. Figure **b)** shows the pressure distribution in the channel superimposed with the velocity vectors of the continuous phase.

The apparent differences between model and experimental data during the necking stage before pinch-off due to the enhanced influence of wall boundary ($C_{b1}$) could be attributed to a) the presence of dynamic wetting conditions in the experiments, whereas the simulations employ a static contact angle and b) the formation of thin liquid film between the droplet and wall in the experiment that is not resolved in the current simulations. However, the drop formation frequency predicted by the model shows a difference of ~1% w.r.t to the experimental data of Glawdel et al. (2012), as shown in **Table.2**, suggesting a good agreement between the model and experiment. Fig.3**b)** shows the intricate features of the overall pressure and continuous phase velocity fields during the formation of the droplet. Features such as the *a)* drop emergence into the channel at 5ms shows an increase in the velocity field due to distortion of the fully developed continuous fluid, *b)* during the filling stage at 15ms, the region between the boundary $C_{b1}$ and the interface experiences significantly higher velocity, *c)* diminishing of the velocity vectors at 30ms just prior to breakup, a significant volume of fluid fills the channel, and finally *d)* the high flow is directed into a region where the droplet pinches off. These critical features are consistent with that observed by the experimental demonstrations of van Stejin et al. (2007) and the 3D numerical work of Soh et al. (2016). Both spatially and temporally, resolving the formation and migration of droplets in microchannel T-junctions using the 3D numerical simulations is highly challenging and computationally expensive. Despite being computationally prohibitive, well-resolved 3D simulations have shown the ability to accurately resolve the lubricating film's dynamics that signify the droplet-wall interactions (Ling et al. (2016)) for a wide range of $Ca$ values. Unlike the 3D direct numerical simulations, the 2D numerical framework adopted in this study is inherently limited to predicting the leakage and corner flow characteristics revealed in the experiments (Korczyk et al. (2019)). Nevertheless, the comparisons illustrate that the 2D approach adopted in this work is able to replicate the essential features of drop formation, viz., the drop length and the formation frequency of the droplet, which are crucial parameters of interest for the present study.

In the following sections, firstly, the hydrodynamic conditions that lead to various flow regimes within the microfluidic T-junction are identified. As a next step, the effect of upscaling is analysed in the context of the droplet morphologies, formation frequencies, transitions, phase maps, and scaling laws by embedding a wide range of JGs within the microfluidic T-junction that is subjected to the hydrodynamic conditions corresponding to the identified flow regimes.

## 4. Results

### 4.1 Identification of flow regimes in the Standard microfluidic-T junction

As described in Section 2.2, the variation of two governing parameters viz., the $Ca$ and $q$ leads to several flow regimes in the present system. Earlier studies have shown that the flow regimes that are present within a microfluidic T-junction can be classified as squeezing, dripping, jetting, and parallel flow where no droplets are formed (Li et al. 2019, Li et al. 2012, Santos and Kwaji, 2010). In addition, the transition between each of these regimes exists that can potentially result in a wide range of flow regimes. Furthermore, the experimental results of Santos and Kwaji (2010) illustrated a snapping regime at very low $Ca$ and with low dispersed phase velocities. However, based on the competition between the forces that are involved within the flow regimes, viz., the surface tension force, shear force on the interface, and the hydrostatic pressure differences on the sides of the emerging droplet, the nature of slugs that are formed differ (Gastecki et al., 2006, Li et al., 2012, Christopher et al., 2008).

Fig.4 shows the blocking ($B_s$), squeezing ($S_s$), dripping ($D_s$), jetting ($J_s$), and parallel flow ($PF_s$) regimes that are identified with the $Ca$ and $q$ values respectively. The subscript 's' denotes that the regimes correspond to the hydrodynamic conditions associated with the standard microfluidic T-junction. In each case, the transient pressure at $P_j$ is plotted since it provides adequate information on the inherent droplet breakup characteristics as adopted previously (Wong et al., 2017, Li et al., 2012). At lower capillary numbers ($Ca \ll 0.032$), two regimes viz., the $B_s$ and the $S_s$ are identified. In Fig.4a) within the images show by **a-e**, the $B_s$ regime is identified with $Ca = 0.0015$ and $q = 2.001$, the dispersed phase enters the main channel that encounters the junction point leading to a pressure increase as shown by the image **a**. Soon after blocking the channel shown by **c**, a slug-like feature evolves with a localized neck that appears closer to the junction, as shown by **d**, leading to pressure build-up at the junction from the point **c**. However, this slug-like feature continues to grow wherein the neck moves away from the junction and shows significant resistance to breaking by growing and therefore, the entire channel cross-section is fully blocked as shown in **e**. The $B_s$ regime described in this study inherits the some characteristics of the snapping regime experimentally described by Santos and Kwaji (2010), except that in the present study except that the pinch-off did not occur at least until t*(s)~0.13s. Experimentally, the work of Arias and Montalur (2020) presented the $B_s$ like regime for small $Ca$ values (as small as 0.6x10$^{-3}$) when analysing the bubble breakup in a microfluidic T-junction. In addition, the 3D numerical results of Li et al. (2014) predicted a long-slug formation that exhibits the same characteristics as that identified in the present study. In their results, Li et al. (2014) observe that such long slugs break with higher wall adhesion forces due to longer wall contact times. The squeezing regime $S_s$, as shown by the images **a-f** within the Fig. 4**b**), exhibits the features such as filling (show by **c**) and necking (shown by the image **d**) that are similar to the $B_s$ regime but eventually the breakup (shown by images **e** and **f**)) is evidenced through the higher pressure in the image **e** relative to that predicted in the image **d**. Although the $S_s$ regime undergoes same sequence in terms of emerging, filling and blocking the channel, it is quite well known that the pressure difference across the slug causes breakup at low $Ca$ such as the squeezing regime (Gastecki et al., 2006, Li et al., 2012, Christopher et al., 2008, Li et al., 2019, De Menech et al., 2008). However, interestingly, this characteristic increase in transient pressure during breakup witnessed in the $S_s$ regime is not evidenced with the $B_s$ regime suggesting that in the $B_s$ regime, the transient pressure that has evolved at the junction was not sufficient to induce a breakup at least until the simulated time $t^*(s)$~0.13s. The dripping regime $(D_s)$, as shown in Fig.4**c**) with $Ca = 0.0322$ and $q = 0.1458$, is shear-dominated, where the breakup occurs when the shear forces overcome the interfacial force.

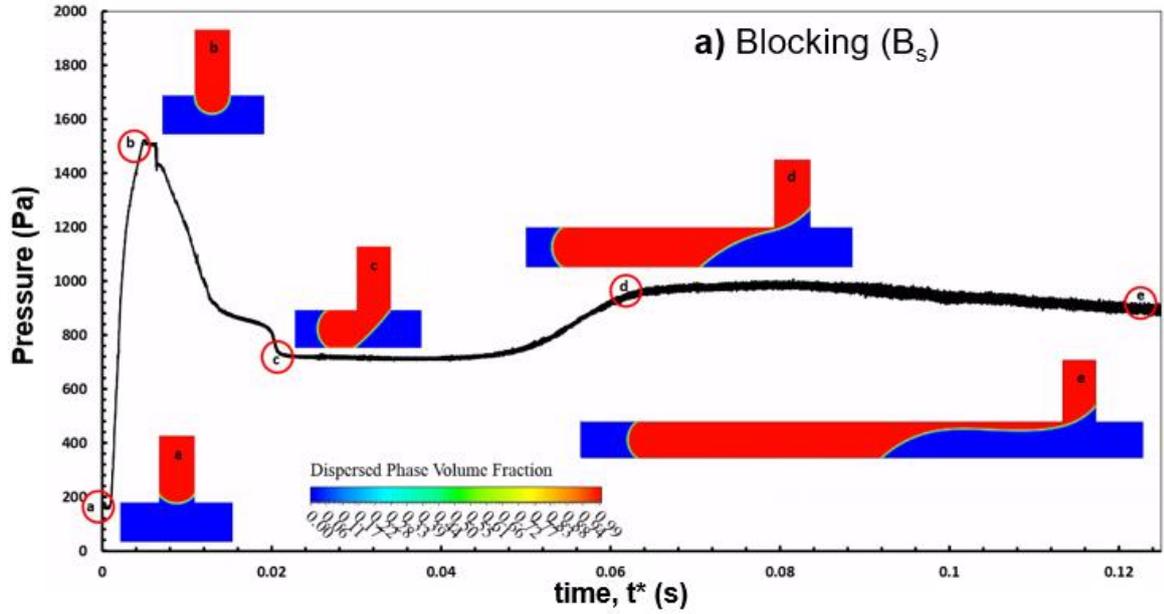

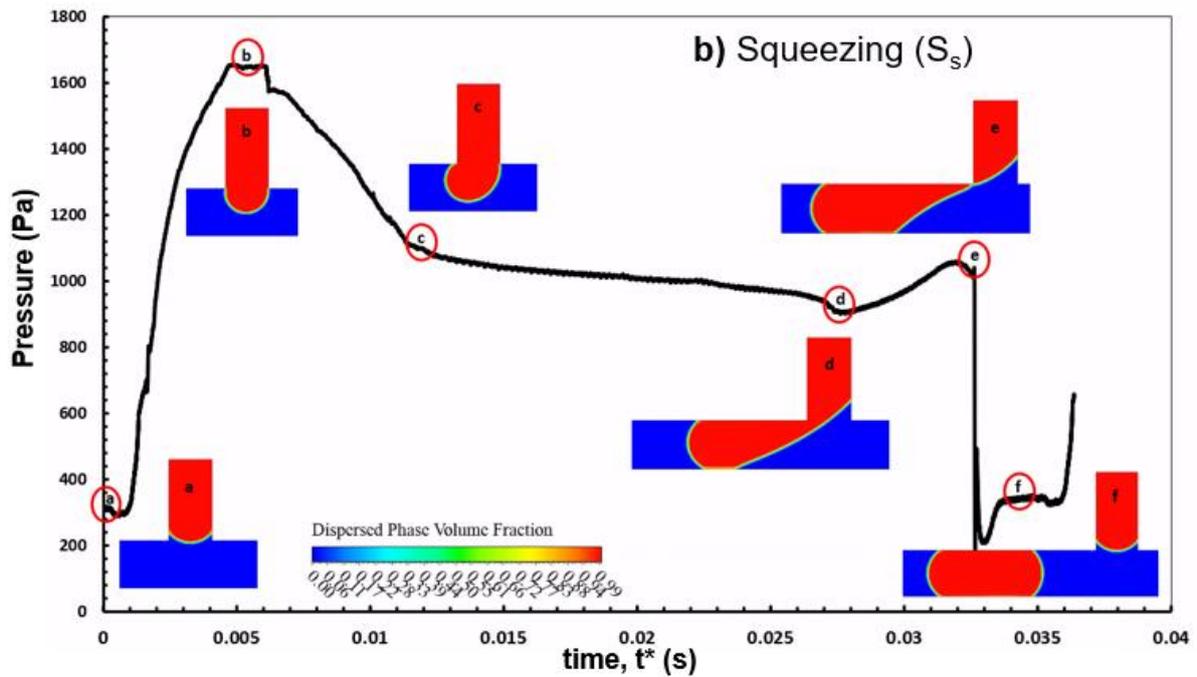

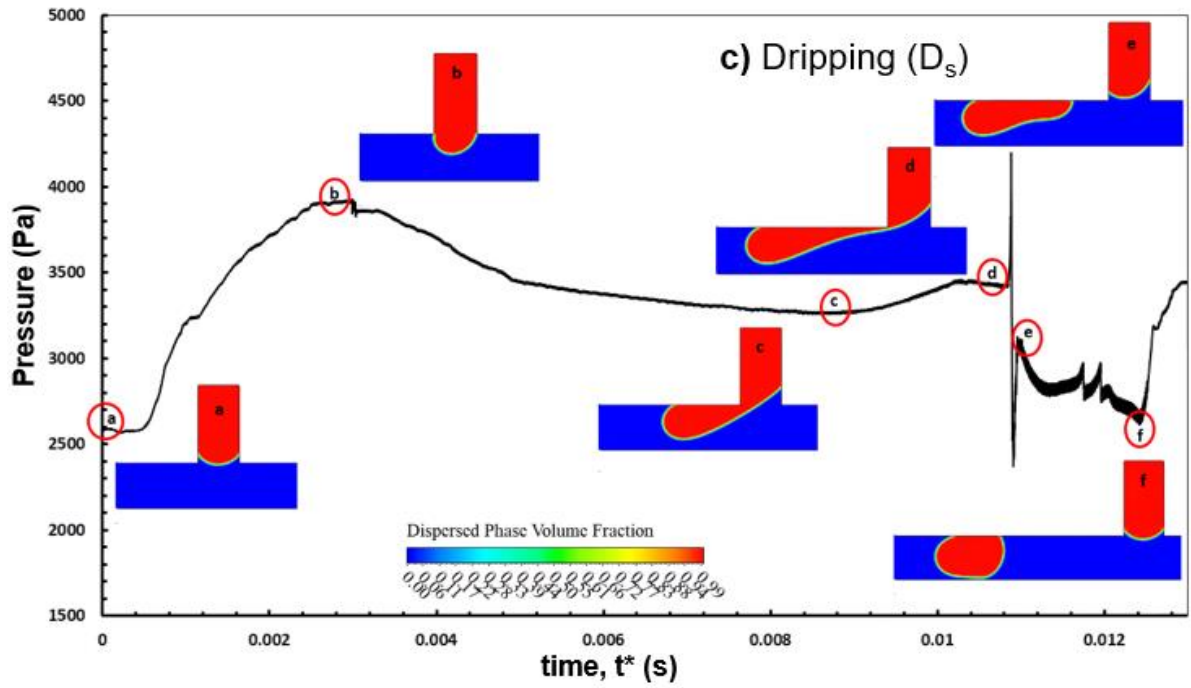

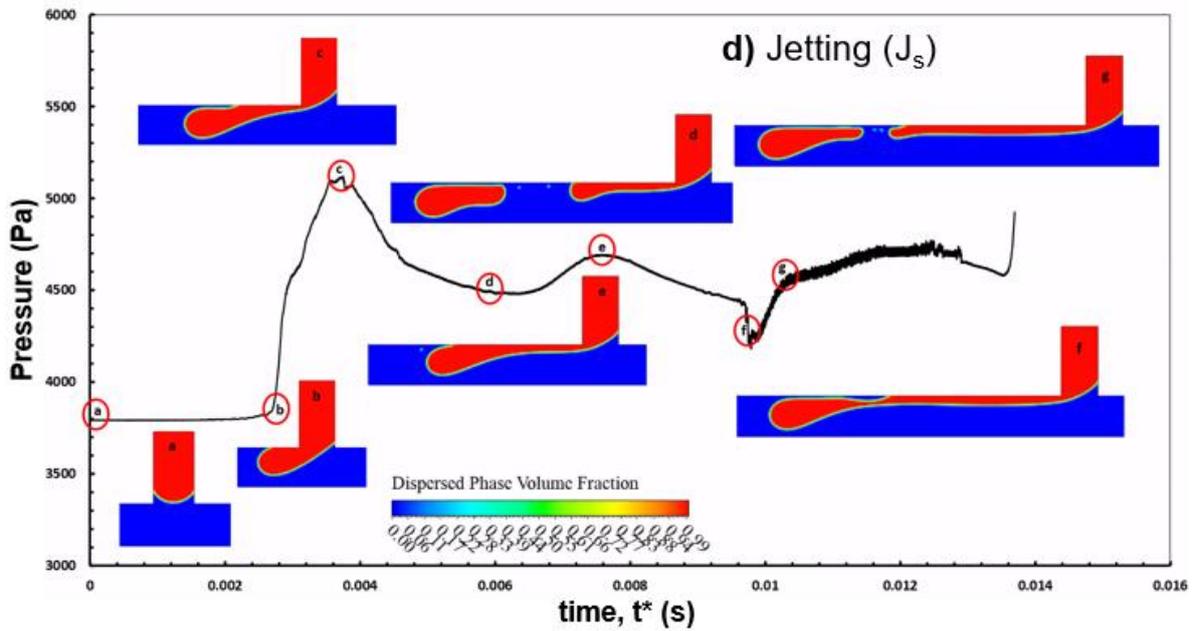

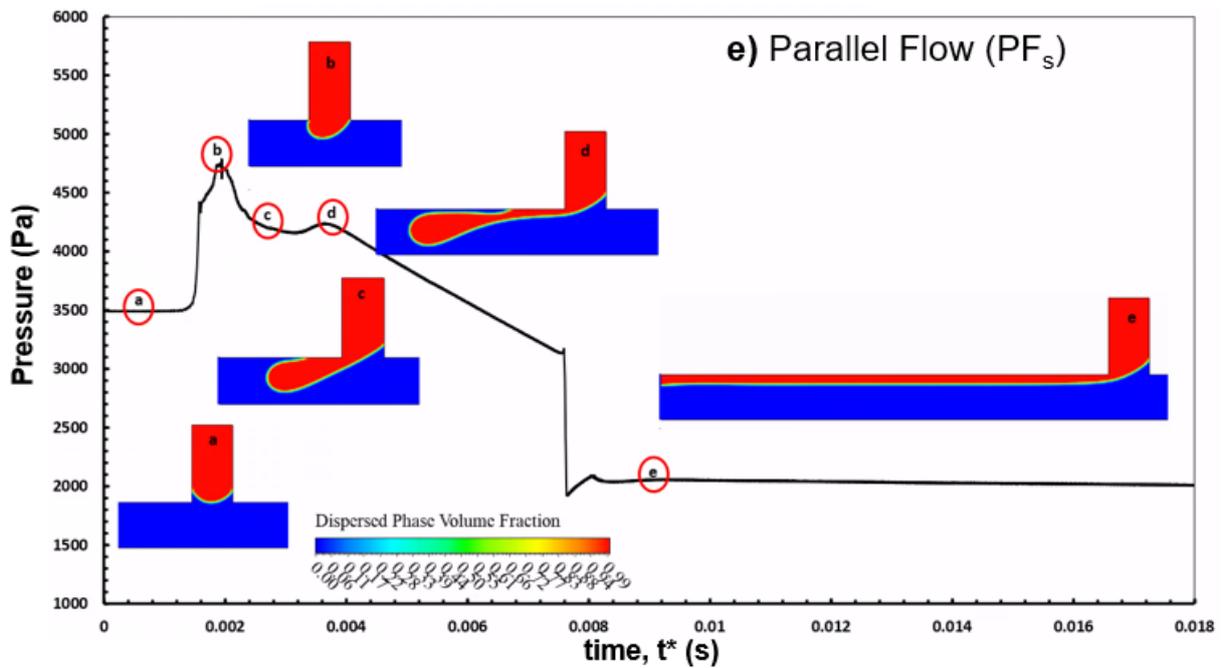

**Fig. 4**. Different flow regimes illustrating several sequences of **a)** Blocking (B$_s$) with $Ca$ = 0.0015, $q$ = 2.001, **b)** Squeezing (S$_s$) with $Ca$ = 0.0043, $q$ = 0.6487, **c)** Dripping (D$_s$) with $Ca$ = 0.0322, $q$ = 0.1458, **d)** Jetting (J$_s$) with $Ca$ = 0.049, $q$ = 0.3296 and **e)** Parallel Flow (PF$_s$) with $Ca$ = 0.0725, $q$ = 0.2777 in the standard T-junction. In each case (**a)-e)**), the transient pressure at the junction probe (P$_j$) is shown alongside the sequence of events described by (a-f). In each case, t*=0s corresponds to the time that the dispersed phase volume fraction first emerges into the continuous phase channel.

Unlike the B$_s$ and the S$_s$ regimes, the droplet does not block the main channel entirely in the case of the D$_s$ regime where the droplet pinch-off is confined to the lower boundary C$_{b2}$ (see Fig.1); therefore, the droplet breakup is accompanied by the continuous fluid that emerges through the gap between the droplet and the upper boundary C$_{b1}$ as seen in the image **d** in Fig.4**c**). With $Ca$ = 0.049 and $q$ = 0.3296 eventually leads to the jetting regime J$_s$ where the dispersed phase fluid shows a thread-like structure. In contrast to the B$_s$ and the D$_s$ regimes where the breakup occurs at the junction, the droplets break up in the J$_s$ regime occurs downstream of the channel as shown by the images **d**, **g** in Fig.4**d**). Finally, the parallel flow regime PF$_s$ is realized with $Ca$ = 0.0725 and $q$ = 0.2777 where the dispersed phase eventually flows parallel to the channel boundary C$_{b2}$ (shown in image **e** within Fig.4**e**)), where no droplets are generated. In addition to the validation shown in Fig.3 and that presented in **Appendix B**, the parameters $\theta$, $\eta$, $q$, and $Ca$ established in this study to characterize the S$_s$, D$_s$, and the J$_s$ regimes, which are in very good agreement with conditions identified for squeezing, dripping, and jetting in a standard T-junction as shown in other previous works (De Menech et al. 2008, Li et al. 2019, Liu and Zhang, 2011). In the context of a gutter that is positioned closer to the junction, as shown in Fig.1, the J$_s$ is representative of the PF$_s$ considering the similarities between the flow behaviour closer to the standard junction as shown by the images **e** and **f** within Fig. 4**d**). Therefore, as a next step, a wide range of junction gutters are introduced into the channel that is subjected to the hydrodynamic conditions pertaining to B$_s$, S$_s$, D$_s$, and the J$_s$ but excluding the PF$_s$.

### 4.2 Influence of junction gutters in the hydrodynamic conditions pertaining to the Bs and the Ss Regimes

Fig.5 shows the effect of drop formation due to junction gutters ranging between $a^*$=0.05 to 1.00 and $b^*$=0.10-0.70 when the microchannel T-junction is subjected to hydrodynamic conditions pertaining to the B$_s$ regime as described previously in Fig. 4**a)**. The introduction of a gutter in the B$_s$ regime promotes the droplet breakup in the junction, unlike that witnessed in the standard T-junction. The slug size appears to be largest for the case with the lowest gutter depth $b^*$=0.10 together with a greater gutter length $a^*$=1.00 analyzed in this study. As shown in Fig.5, although both $a^*$ and $b^*$ influence the length of the droplet $L_d$, however, the gutter depth $b^*$ appears to predominantly influence the droplet size compared to the gutter length $a^*$ with the hydrodynamic conditions pertaining to the B$_s$ regime. In addition to the droplet size, the variation of gutter dimensions has a broader consequence in terms of the morphological characteristics pertaining to the breakup, as demonstrated in Fig.6 and Fig.7.

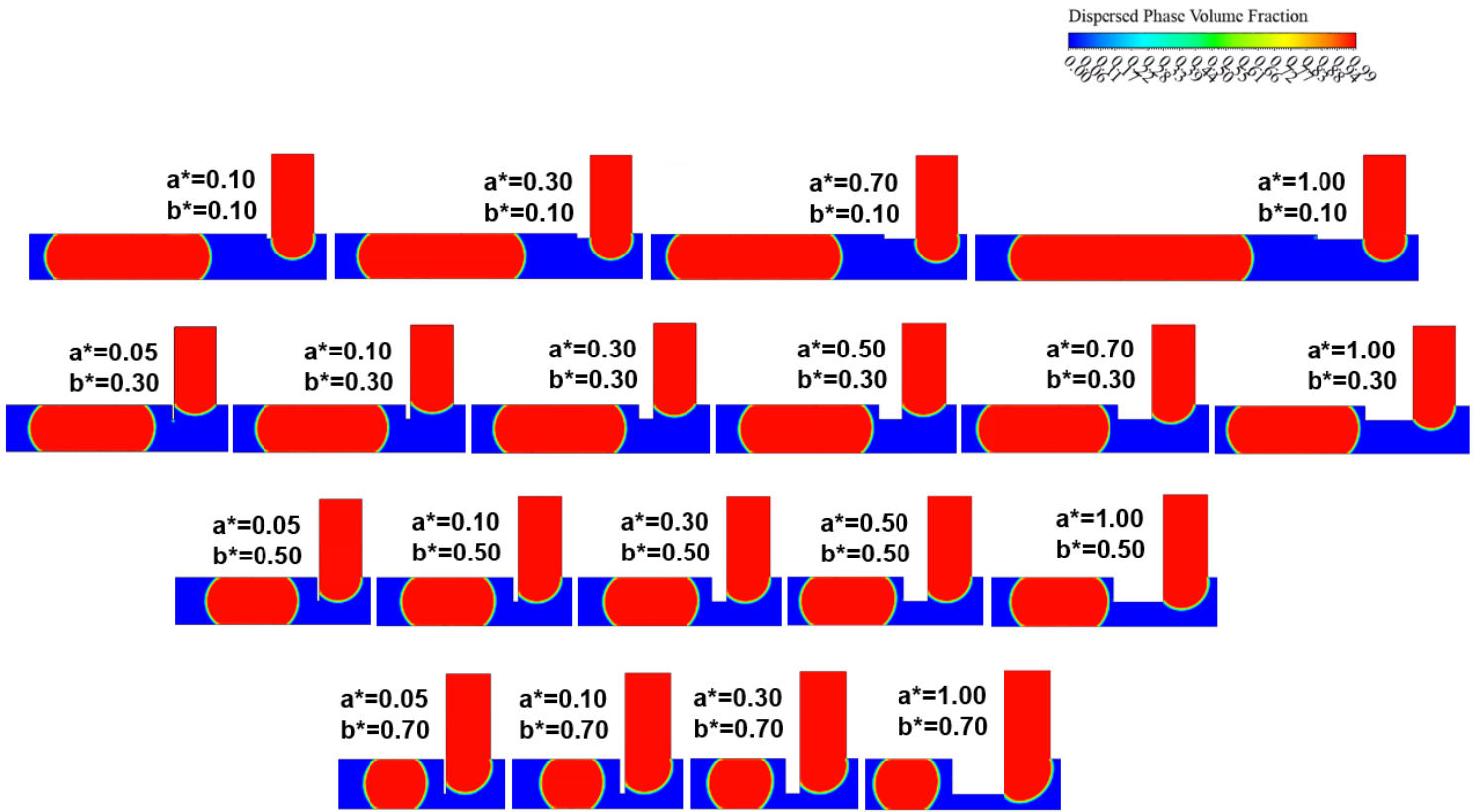

**Fig. 5**. Volume fraction profiles showing the droplet formation with $Ca$ = 0.0015, $q$ = 2.001 for several junction gutter combinations ranging from $a^*$=0.05-1.00, and $b^*$=0.10-0.70.

The influence of droplet breakup for a fixed value of $b^*$=0.10 but with varying $a^*$=0.10-1.00 when subjected to the hydrodynamic conditions of B$_s$ is shown in Fig 6. For each gutter configuration, the sub-figure **a**) shows the necking just prior to slug pinch-off whilst the figure **b**) shows the incipience of the slug breakup.

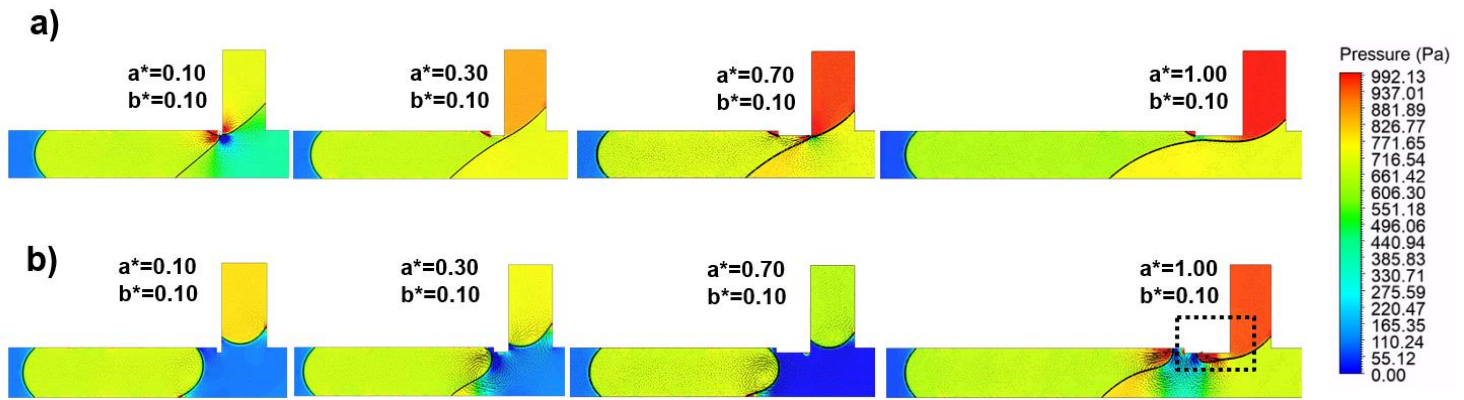

**Fig. 6**. Pressure distribution for several junction gutter combinations with $Ca$ = 0.0015, $q$ = 2.001, $b^*$=0.1 (fixed), and $a^*$=0.1-1 (varied). In each case, the figure **a)** shows the necking behaviour prior to drop detachment and figure **b)** shows the droplet detachment corresponding to the figure on the top panel.

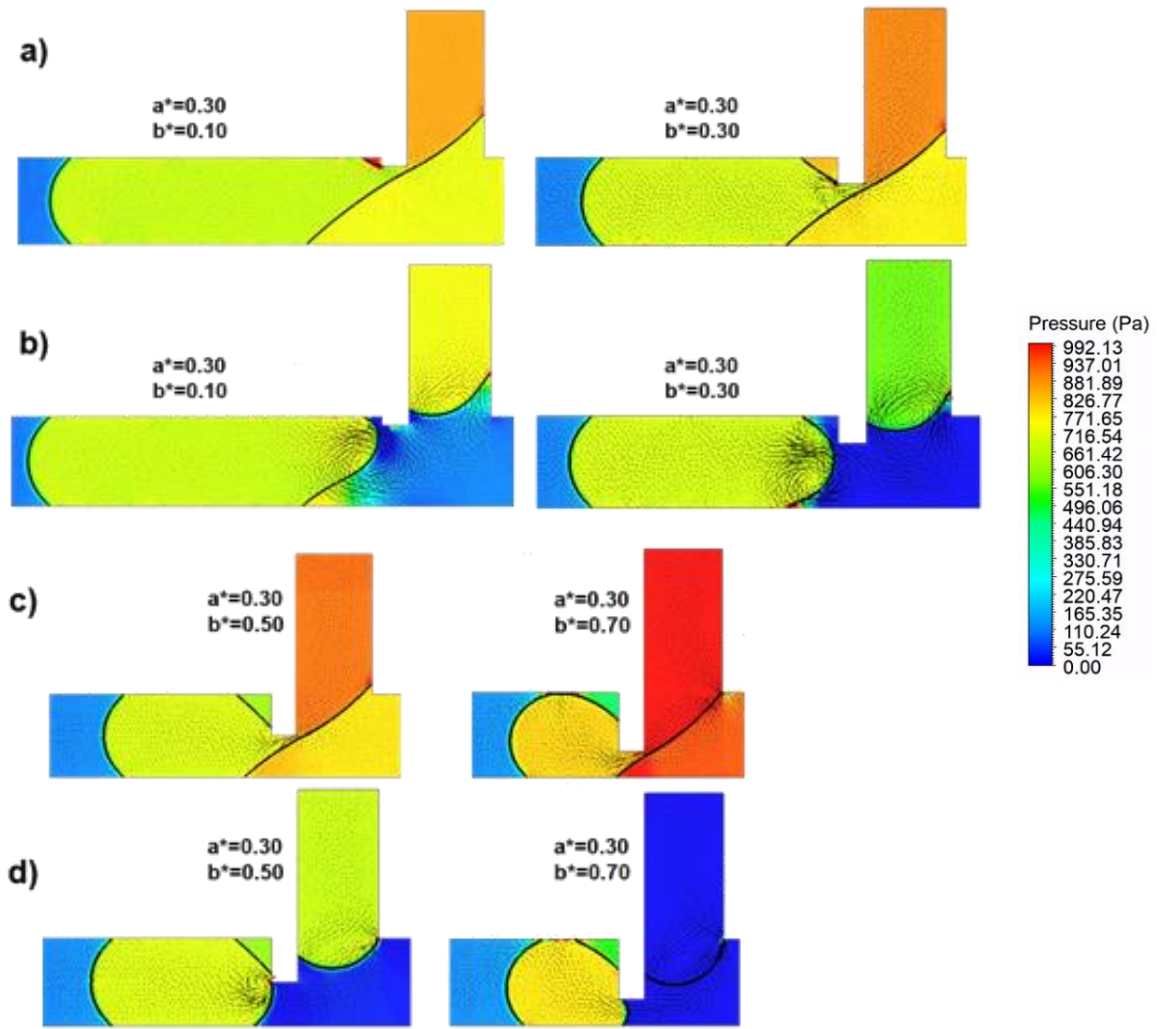

**Fig. 7**. Pressure distribution for several junction gutter combinations with $Ca$ = 0.0015, $q$ = 2.001 $a^*$=0.30 (fixed) and $b^*$=0.10-0.70 (varied). In each case, the figures **a)**, **b** shows the necking behaviour prior to drop detachment and figures **c)**, **d)** show the droplet detachment corresponding to the figure on the top panel.

For $a^*$<1.00, the necking process is at the junction, and the images (at the bottom) show that the interface recedes into the main channel after the slug is formed. Interestingly, the necking and breakup sequences described in all cases with $a^*$<1.00, and $b^*$=0.10 are reminiscent of the $S_s$ regime, although the channel is subjected to hydrodynamic conditions pertaining to the $B_s$ regime. However, with $a^*$=1.00 and $b^*$=0.10, the slug grows into the main channel with a thin liquid thread but eventually snaps at a distance from the junction, unlike in the other cases. Eventually, the thin liquid tail tends to retract into the T-Junction as shown by the dotted square box in the sub-figure **b)** $a^*$=1.00 and $b^*$=0.10. It can be observed that the process of (i) tail thinning away from the junction point and (ii) the breakup of the large slug closely resembles the snapping slug described by Santos and Kwaji (2010). The effect of variation in gutter depth ($b^*$) for fixed gutter length ($a^*$) is shown in Fig.7. The post-breakup images shown by the cases with $b^*$=0.50 and 0.70 suggest that an increase in the gutter depth significantly reduces the slug size, and the slug evolves eventually by squeezing in between the gutter and channel walls. It is interesting to note that, for every gutter topology besides the small gutter case ($a^*$=0.10 and $b^*$=0.10) shown in both Fig.6 and Fig.7, the dispersed phase channel experiences lower pressure distribution during post-breakup than compared to the necking stage.

Fig.8**a)** presents the variation of gutter length ($a^*$=0.05-1.00) with fixed gutter depth ($b^*$=0.30) on drop formation when the microfluidic T-junction is subjected to the hydrodynamic conditions pertaining to the $S_s$ Regime (see Fig. 4**b)**). In each case, the images show necking (right column), breakup (middle column) and location of the droplet in the channel (left column) extracted at the same instance of time ($t^*$=0.03215) for all cases to assess the speed of droplet traverse in the main channel. With small values of $a^*$=0.05 and 0.10, the necking and pinch-off responses shown in Fig.8**a)** no longer exhibit the characteristics of the $S_s$ regime, i.e., during necking or pinch-off, the droplet does not block the channel by adhering to the channel boundary ($C_{b1}$) instead, characteristics like that of a dripping regime $D_s$ (see Fig.4**c)**) is observed. However, for cases with $a^* \geq 0.30$, the necking and pinch-off characteristics of $S_s$ re-appear and are well preserved. To distinguish the nature of breakup shown by $a^*$<0.30 against that observed with $a^* \geq 0.30$, the transient pressure evolution at the junction point ($P_j$) for two different cases is presented in Fig 8**b)**. For $a^*$=0.05, the incipience of pinch-off indicated by blue squares shows high-pressure peaks, which is a characteristic of the $D_s$ regime observed in Fig.4**c)** (shown by the images **d**) and **e**)). With $a^*$=0.50, although the pressure at the junction increases during pinch-off (shown by green circles), the peak observed at the incipience of droplet pinch-off is markedly different to $a^*$=0.05 but shows similarities to the pressure profile of the $S_s$ regime. This suggests that the choice of gutter configurations in the microchannel can influence transitions at least during the necking phase and drop detachment at low $Ca$. For all cases shown in Fig.8**a)** (left column), the results predict that the speed of the evolved droplet in the main channel appears to increase with incremental values of $a^*$.

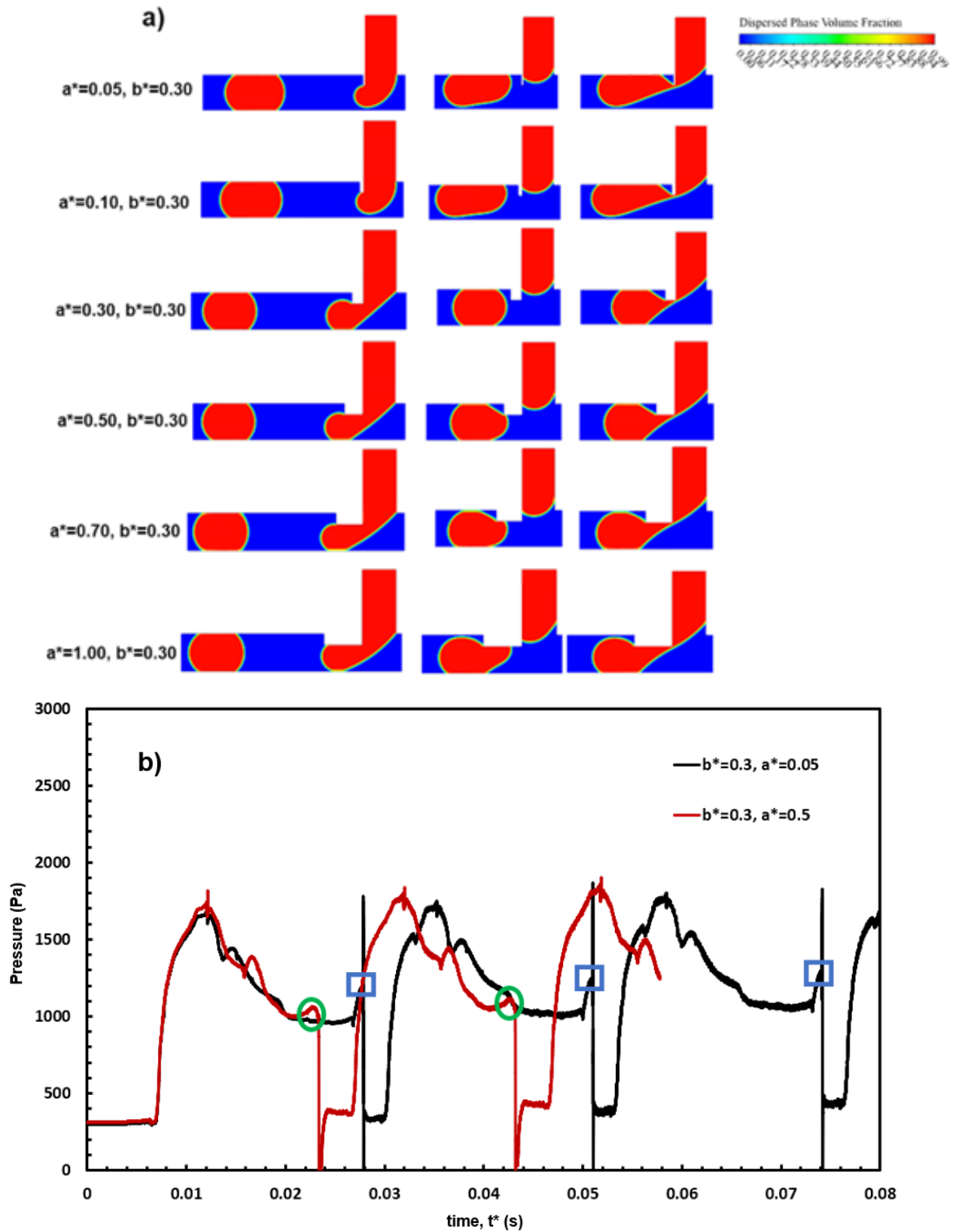

**Fig. 8**. **a)** Volume fraction profiles showing the droplet formation with $Ca$=0.0043, $q$=0.6487 for several junction gutter combinations with $b^*$=0.30 (fixed) and $a^*$=0.05-1.0 (varied). For each case (from right to left), the necking, droplet detachment, and the location of the droplet in the channel extracted at $t^*$=0.03125s; where $t^*$=0 is identical to that presented in **Fig. 4.b)**. Figure **b)** shows the transient pressure evolution at the junction probe ($P_j$) for two cases where $b^*$=0.3 and $a^*$=0.05 and 0.5. The droplet formation times are indicated by green circles (for $a^*$=0.05) and blue squares (for $a^*$=0.5).

The effect of gutter dimensions on the size of the droplet ($L_D{}^* = \frac{L_D}{W_c}$) and the droplet formation frequency (*f*) is summarized in Fig.9 when the channel is subjected to conditions corresponding to the Bₛ (show by symbols with solid lines) and the Sₛ (shown by symbols with dashed lines). For cases pertaining to the Bₛ, except for $b^*$=0.10, both $L_D{}^*$ and *f* show no noticeable variation w.r.t to the gutter length a*. However, the variation of $a^*$ has a significant impact when $b^*$ is small, viz., with 0.10, which is supported by the images of drop evolution shown in Fig.6). The behavior is similar with the conditions of Sₛ where, for most cases, the change in $a^*$ has negligible influence on $L_D{}^*$ and *f*. However, the presence of a transition shown for $a^*$<0.30 to $a^*$≥0.30 suggests that the choice of $a^*$ can influence a transition in the system, as shown in Fig.8). For all cases investigated, the predictions suggest that increasing the gutter depth $b^*$ for all values of $a^*$ significantly reduces the size together with increasing the frequency of the droplet formation when the channel is subjected to the Bₛ as well as the Sₛ conditions.

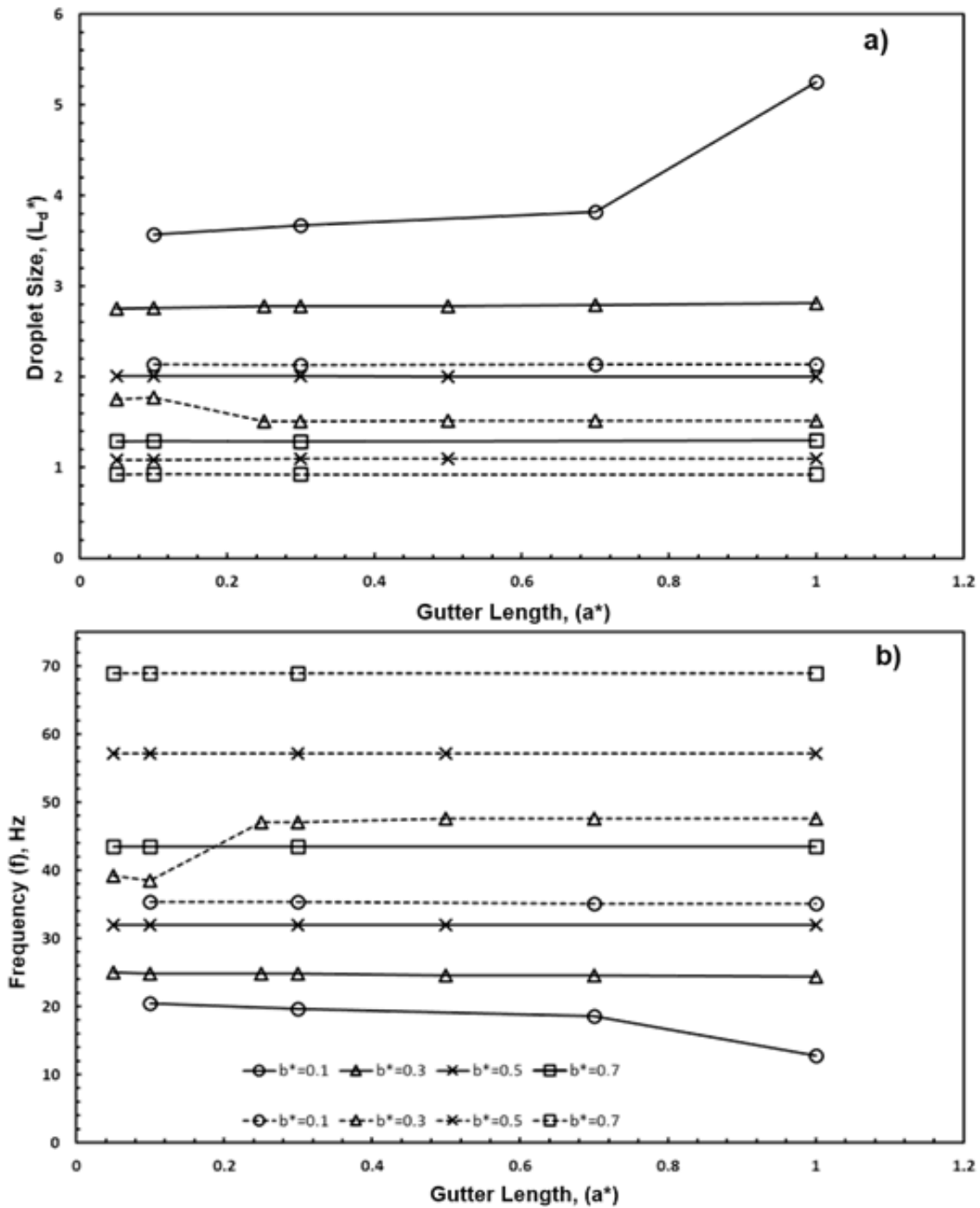

**Fig. 9. a)** shows the droplet size ($L_D{}^* = \frac{L_D}{W_C}$) and **b)** the droplet formation frequency ($f$) against the gutter length ($a^*$) for several values of gutter depth ($b^*$=0.10-0.70). In each case, the symbols with solid lines and dashed lines correspond to hydrodynamic conditions $Ca$=0.0015, $q$ =2.001 and $Ca$=0.0043, $q$=0.6487 respectively.

Several researchers have proposed scaling laws for predicting the size of the slug ($L_D{}^*$) for over a range of $Ca$ values within the (i) squeezing (Garstecki et. al., (2006)) and (ii) transition from squeezing to dripping regimes (Xu et al., (2008)) for standard T-junctions. More recently, the work of Li et al. (2019) proposed a scaling relationship within the squeezing regime for T-junction microchannels with rectangular ribs, which suggested that the resulting agreement between the numerical data and derived scaling law could be as close as ~15%. However, their study was limited to rib depth of up to 50%. In the current work, the general equation proposed by Li et al. (2019) is adopted in Eq. (18), but as a function of velocity ratios ($q$) of the continuous and dispersed phase fluids, as follows:

$$L_D{}^* = \frac{L_D}{W_c} = (1 - b^*)\gamma + (1 - b^*)^2\beta q \tag{18}$$

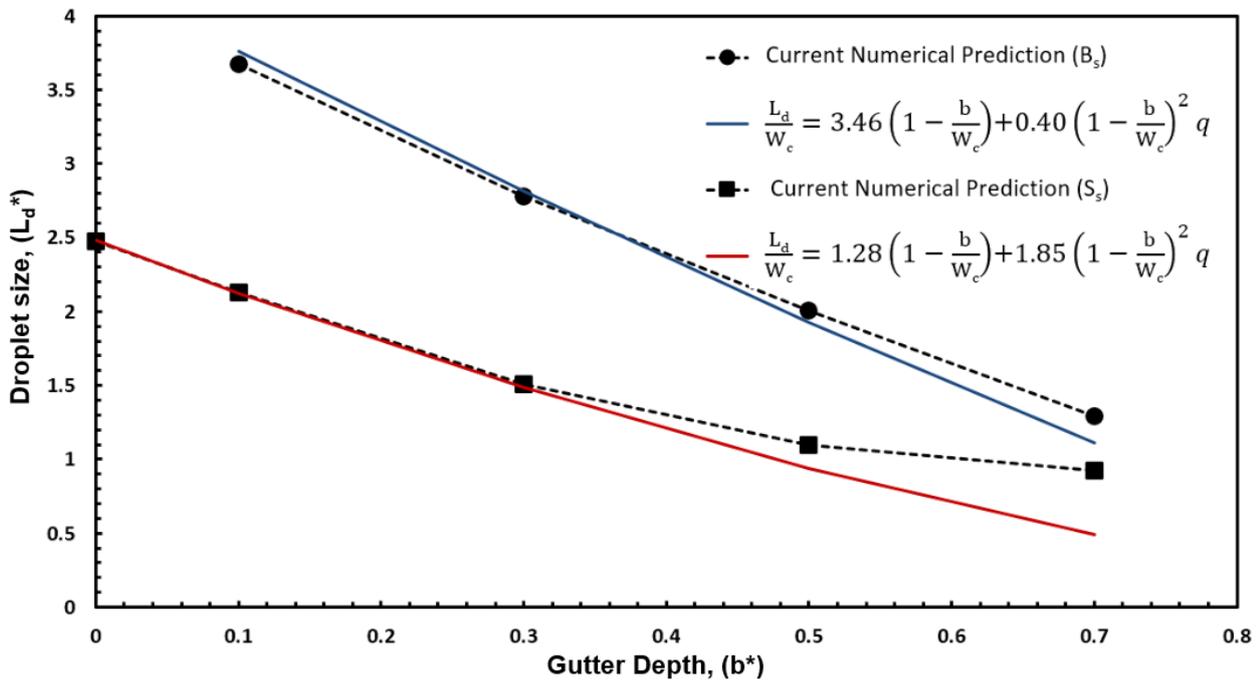

**Fig. 10.** A comparison of the normalized droplet size predicted by current numerical work against the droplet size obtained based on the Eq. (18) for hydrodynamic conditions that correspond to (Bₛ) regime (See **Fig. 4a)**) with $Ca$=0.0015, $q$=2.001, and (Sₛ) regime (See **Fig. 4 b)**) with $Ca$=0.0043, $q$=0.6487 respectively.

In Eq.18, $\gamma$ and $\beta$ are fitting parameters. To verify the predictions of the numerical model, the resulting slug size is compared against theory (Eq. 18) for $a^*$=0.30 but tested for a wide range of gutter depths ($b^*$). The estimated using Eq.18 for both Bₛ and Sₛ conditions resulted in Eq. 19 and Eq. 20.

$$L_D{}^* = 3.46\ (1 - b^*) + 0.40\ (1 - b^*)^2 q \tag{19}$$

$$L_D{}^* = 1.28\ (1 - b^*) + 1.85\ (1 - b^*)^2 q \qquad (20)$$

When the channel is subjected to the Bs conditions, the $Ca$=0.0015 (small) and therefore, the overall agreement with the theory (Eq.18) is <~4% for all cases except for $b^*$=0.70 where the deviation is up to ~16% as shown in Fig. 10. With hydrodynamic conditions pertaining to Ss, the resulting increases to $Ca$=0.0043, and consequently, the nonlinearities associated with the breakup process significantly increases, thereby resulting in deviations of ~16% with $b^*$=0.50 and as high as 88% for $b^*$=0.70. Nevertheless, the scaling law described in Eq. 18 agrees well for $b^* \leq 0.50$ and $b^*$<0.5 for conditions pertaining to Bs and Ss, respectively. To further analyse the consequence of nonlinearities associated with the breakup process, the channel is subjected to Ds and Js conditions with several junction gutters.

### 4.3 Influence of junction gutters in the hydrodynamic conditions pertaining to the Ds and the Js Regimes

Fig.11**a)** presents a phase space over a wide range of gutter depth ($b^*$) and gutter length ($a^*$) that occur at $Ca$=0.0322, $q$=0.1458, which correspond to the conditions of the Ds regime. Three distinct droplet morphologies in terms of adherence to channel walls of the T-junction are predicted for over a wide range of ($a^*$,$b^*$) in the flow map as shown in Fig.11**b)**. The limits marked by the solid red and blue lines in Fig.11**a)** present a clear transition between droplets' adherence to the channel boundaries.

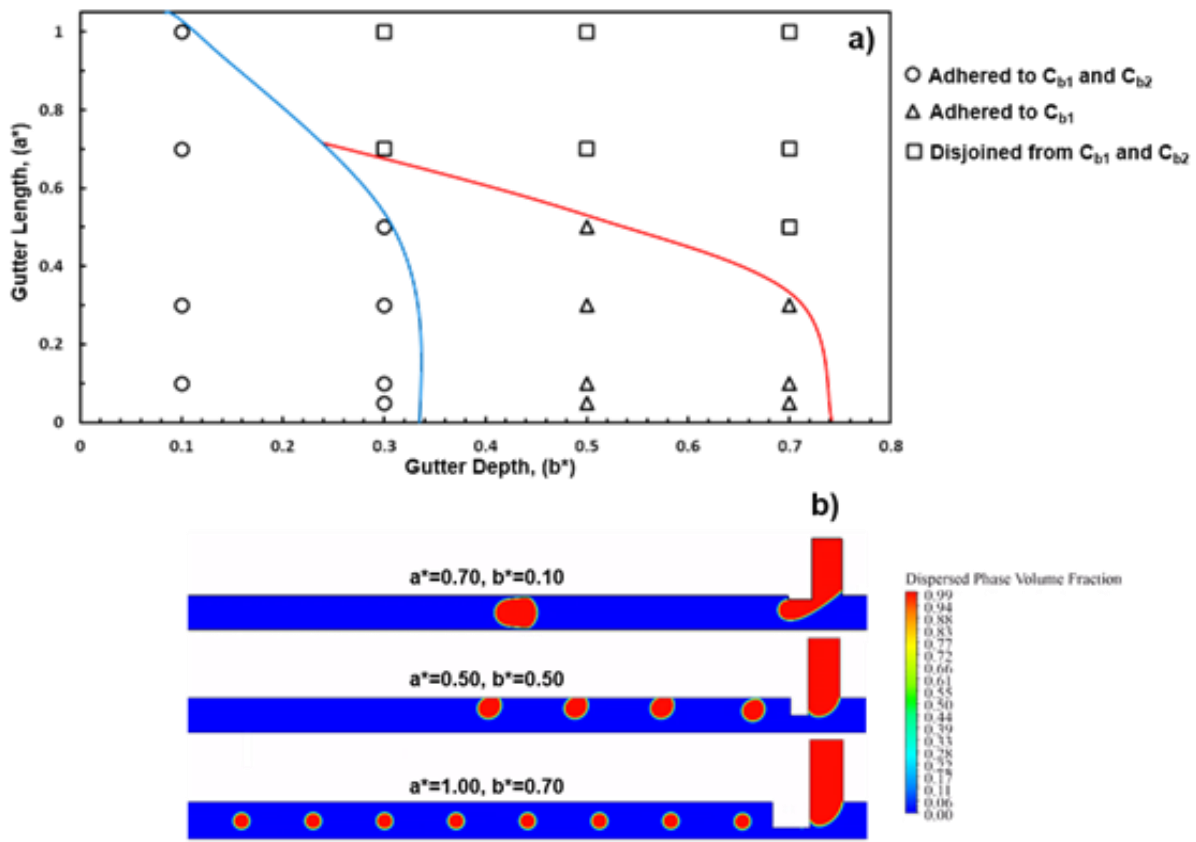

**Fig. 11**. **a)** flow regimes in ($a^*$, $b^*$) marked by solid blue and red lines that indicate different transitions w.r.t droplets adherence to the channel boundary, **b)** shows the typical morphologies of the evolved droplets described by the flow regime map. In all the cases, the hydrodynamic conditions are fixed with $Ca$=0.0322, $q$=0.1458 that correspond to D_s regime as shown in **Fig.4 c)**.

For most cases when $b^*$≤0.30, the droplet formation and evolution processes are much like the dripping regime wherein the drops adhere to both the upper ($C_{b2}$) and lower ($C_{b1}$) channels, indicated by circles in the map. However, with $b^*$=0.30 and for $a^*$>0.60, a transition appears where the evolved drops are much smaller and are unbounded to either of the channel boundaries indicated by open squares. When $b^*$≥0.5, another transition appears where the drops adhere to the boundary $C_{b2}$ as shown by open triangles. This regime (shown by open triangles) diminishes with the increase in $b^*$ and for increasing values of $a^*$ where the unbounded drop regime starts to predominate in the map. Like the earlier conditions of $B_s$ and $S_s$, for any given configuration of the gutter, the droplet spacing ($\lambda$), (see Fig.1), predicted by the numerical model between any two droplets with the $D_s$ conditions, are identical. Consequently, all droplet shapes and drop formation frequencies predicted are identical for every subsequent droplet formed after the first drop, which reinforces that the gutters exhibit strong potential to produce monodisperse drops for conditions pertaining to $D_s$. Fig.12**a)** and Fig.12**b)** complement the information from the regime map by showing the variation in $L_D^*$ and $f$ due to $a^*$. Unlike the $S_s$ and the $B_s$ conditions, as shown in Fig.9, although negligible, the size of drops shows some noticeable variations with change in $a^*$ due to a plethora of morphological transitions that are evidenced in the ($a^*$, $b^*$) phase map shown in Fig.11**a)**. With the increase in $b^*$, that the drop formation frequency can significantly increase, as shown in Fig.12**b**), thereby suggesting that the inclusion of gutters can substantially promote upscaling correspondingly in the $D_s$ regime.

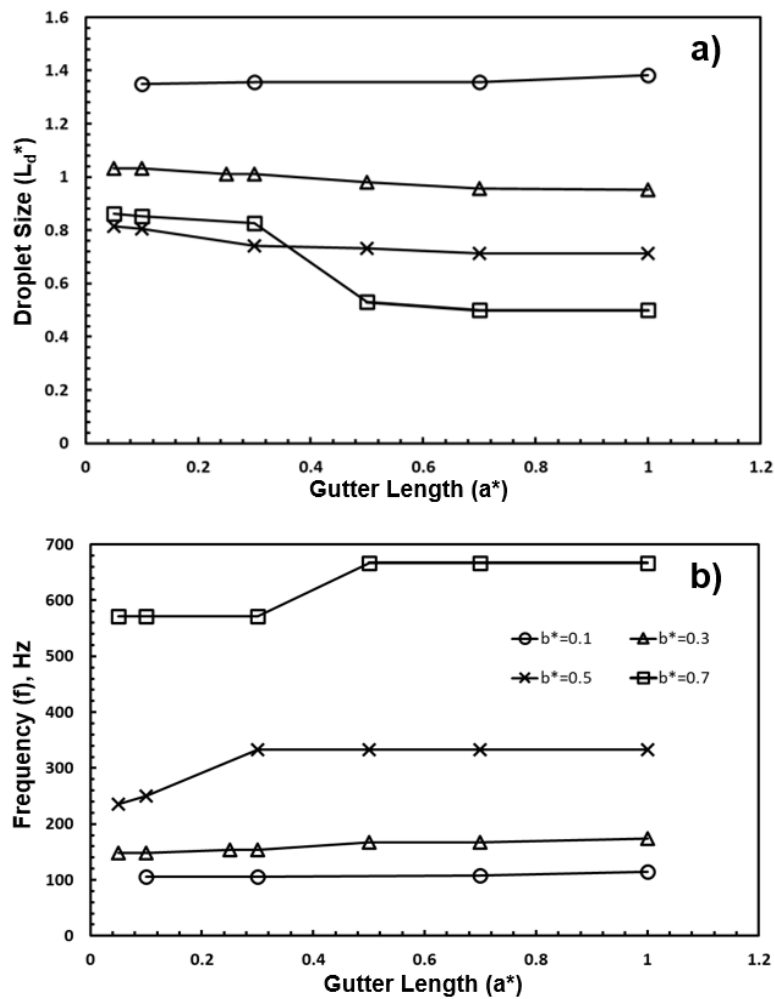

**Fig. 12. a)** shows the droplet size $(L_D{}^* = \frac{L_D}{W_c})$ and **b)** the droplet formation frequency (f) against gutter length (a*) for several values of gutter depth (b* =0.10-0.70) for the hydrodynamic conditions with $Ca$=0.0322, $q$=0.1458 that correspond to D$_s$ regime as shown in **Fig. 4 (c)**.

When the microchannel T-junction is subjected to the J$_s$ regime with $Ca$=0.049, $q$=0.3296, the gutters alter the flow pattern more drastically, as shown by the ($a^*$, $b^*$) flow regime map in Fig. 12**a)**. The attributes of the J$_s$ regime as shown in Fig.4**d)** are no longer preserved; instead, with $b^*$<0.40 and $a^*$<0.4, a uniform dripping regime emerges. The uniform dripping regime inherits the features of the D$_s$ regime and is shown by filled circles in the phase map; its boundaries are defined by the solid red lines. One such formation of droplets in the uniform dripping regime is evidenced in Fig.12**b)** with gutter dimensions $a^*$=0.30 and $b^*$=0.40; together, the transient pressure at the junction (P$_j$) shown by the solid green line suggests that the once the junction pressure stabilizes, the droplets continue to evolve with uniform droplet spacing ($\lambda$), the droplets are monodisperse, and the frequency of formation (f) of the droplets are uniform and stable.

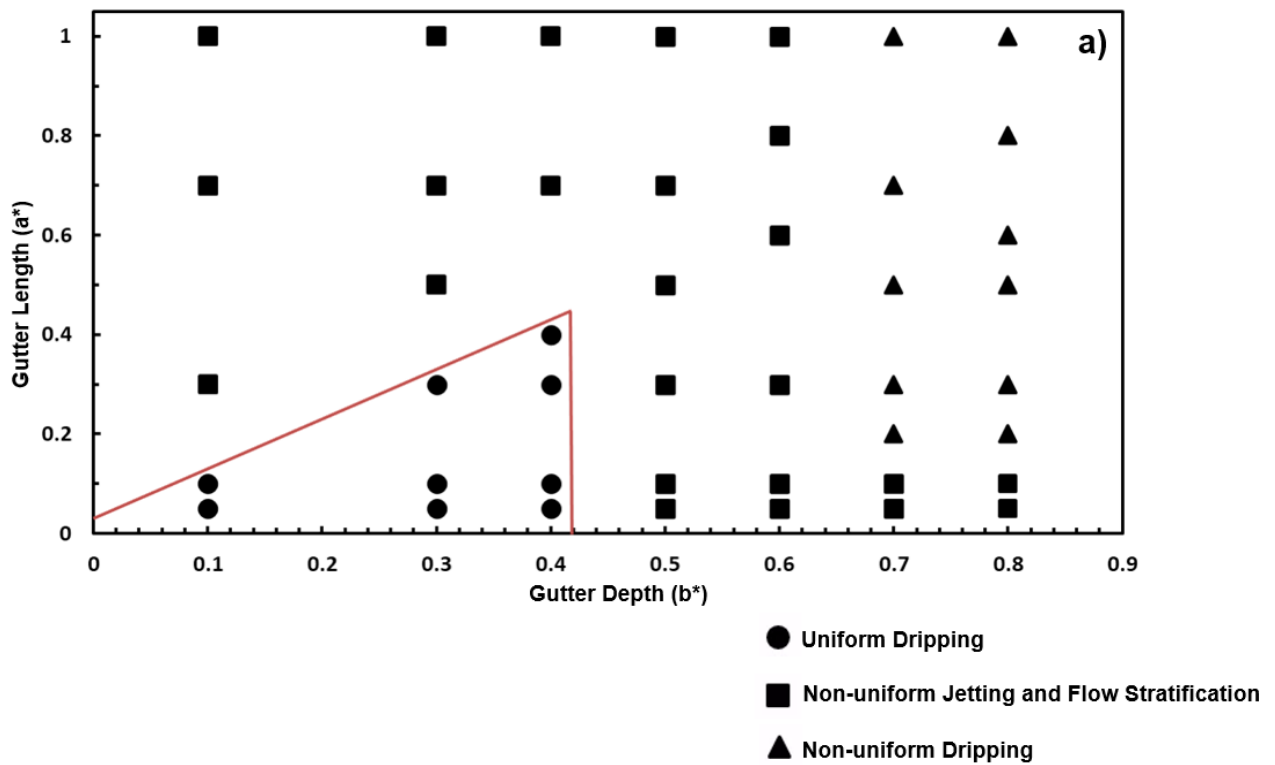

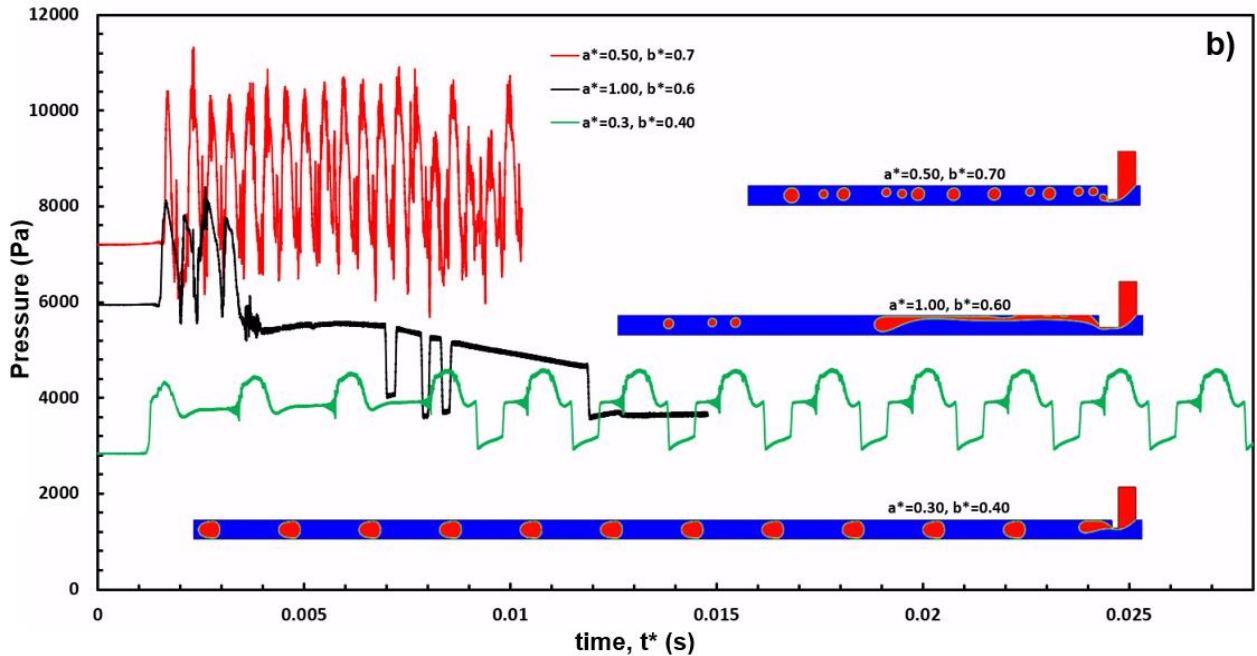

**Fig. 13**. **a)** flow regimes in ($a^*$, $b^*$) with droplets generated through three characteristic modes predicted viz., the uniform dripping, the non-uniform dripping that proceeds to parallel flow, and non-uniform dripping. The solid red lines indicate the boundary of the uniform dripping region. **b)** shows the transient pressure evolution at the junction probe (P$_j$) with different gutter configurations that describe the three regimes and morphologies of the evolved droplets. In all cases, the hydrodynamic conditions are fixed with $Ca$=0.049, $q$=0.3296 which correspond to Js regime as shown in **Fig. 4 d)**.

Crossing the threshold of the uniform dripping leads to the onset of non-uniform jetting (marked by filled squares in Fig. 12**a)**). One instance of this regime is shown by $a^*$=1.00, $b^*$=0.60 in Fig. 12**b)**; the corresponding transient pressure evolution at the junction is shown by the solid black line. This regime shows attributes of dripping at the initial stages, wherein some drops begin to emerge but with non-uniform spacing and size. However, eventually, the dispersed phase liquid continues to evolve with characteristics similar to a parallel flow regime (PF$_s$) as previously shown in Fig.4**e)**. For values of $b^*$≥0.70, the parallel flow characteristics are inhibited, but the non-uniform drips continue to evolve in the channel; this regime is shown by filled triangles in regime map in Fig.12**a)** and is shown by $a^*$=0.50 and $b^*$=0.70 with the corresponding transient pressure at the junction shown by the solid red line in Fig. 12**b)**. Unlike the uniform dripping regime, both the non-uniform jetting and the non-uniform dripping tend to deteriorate the monodispersity in drop formation significantly and tend to become unfavorable.

## 5. Conclusions

In this study, the effect of junction gutters on drop formation in a microchannel T-junction was numerically investigated. The numerical model was comprehensively assessed in the form of a grid verification test, evaluating the choice of interface capturing methods such as VOF and CLSVOF and identifying the rationale behind choosing the VOF method and validated with the experimental data of Glawdel et al. (2012) for a standard microchannel T-Junction. The hydrodynamic conditions leading to flow regimes such as the blocking ($B_s$), squeezing ($S_s$), dripping ($D_s$), jetting ($J_s$) and parallel flow ($PF_s$) that are characterized by the capillary number ($Ca$) and velocity ratio ($q$) were identified for the standard channel. An extensive range of Junction Gutters (JGs) was then embedded onto the junction of the standard microchannel, with gutter lengths $a^*$, and depth $b^*$, ranging from 0.05-1.00 and 0.10 and 0.80, respectively. With the introduction of the JGs under the same hydrodynamic conditions pertaining to regimes mentioned above, the following findings are summarized.

(i)  In the hydrodynamic conditions of $B_s$ and $S_s$ where the $Ca$ and flow velocities are small, gutter depth $b^*$ significantly influences the drop size and frequency of formation of droplets than the gutter length, $a^*$. However, $a^*$ tends to promote transition by invoking changes to the morphology of the breakup of drops for small values of $a^*$. In these regimes, the theoretical scaling law for predicting the size of the droplet with the gutters appears to strongly depend on $b^*$ and matches reasonably well with the numerical predictions for $b^* \leq 0.50$ for both the regimes.

(ii)  With the presence of JGs in the Ds regime, the results suggest the presence of three distinct droplet morphologies of drops which are detailed by the ($a^*, b^*$) phase map in terms of their nature of adherence to the channel boundary of the T-junction. Unlike in the $B_s$ and the $S_s$ conditions, the size and formation frequency of the drops show noticeable variations with $a^*$ when the channel is subjected to $D_s$ conditions. However, when the channel is subjected to $J_s$ conditions, flow transitions such as uniform dripping, non-uniform jetting, and non-uniform jetting occur that are presented through the ($a^*, b^*$) flow map, which details both the favourable and unfavourable topologies of gutters.

(iii)  For the range of flow regimes identified through $Ca$ and $q$ within the standard T-junction, the JGs tend to influence the drop generation rates by promoting the upscaling favourably. Nevertheless, a careful selection of JGs in the hydrodynamic conditions of the $J_s$ regime is vital to foster monodisperse drop generation by transitioning into a uniform dripping regime.

Further numerical and experimental investigations are necessary to underpin (i) the optimal shape of JGs for drop formation, (ii) the influence of wall wettability between JGs and channel boundaries, (iii) critical conditions that can alter flow transition, and (iv) modified scaling laws with gutters that can predict the size of droplets during transitions.

**Acknowledgements**

This work was supported by ANSYS Academic Research Partnership Grant.

# Declaration of Competing Interest

The author declares that there is no competing financial interests or personal relationships that could have appeared to have influenced the work reported in this paper.

# References

1. Abate, A. R., & Weitz, D. A. (2011). Air-bubble-triggered drop formation in microfluidics. Lab on a Chip, 11(10), 1713. https://doi.org/10.1039/c1lc20108e.

2. Arias, S. (2020). Comparison of Two Gas Injection Methods for Generating Bubbles in a T-junction. In Microgravity Science and Technology (Vol. 32, Issue 4, pp. 703–713). Springer Science and Business Media LLC. https://doi.org/10.1007/s12217-020-09790-3

3. Arias, S., & Montlaur, A. (2020). Numerical and Experimental Study of the Squeezing-to-Dripping Transition in a T-Junction. In Microgravity Science and Technology (Vol. 32, Issue 4, pp. 687–697). Springer Science and Business Media LLC. https://doi.org/10.1007/s12217-020-09794-z

4. Bashir, S., Rees, J. M., & Zimmerman, W. B. (2011). Simulations of microfluidic droplet formation using the two-phase level set method. Chemical Engineering Science, 66(20), 4733–4741. https://doi.org/10.1016/j.ces.2011.06.034

5. Basiri, A., Heidari, A., Nadi, M.F., Fallahy, M.T.P., Nezamabadi, S.S., Sedighi, M., Saghazadeh, A., Rezaei, N., 2020. Microfluidic devices for detection of RNA viruses. Rev Med Virol 31, 1–11. https://doi.org/10.1002/rmv.2154.

6. Brackbill, J. U., Kothe, D. B., & Zemach, C. (1992). A continuum method for modeling surface tension. Journal of Computational Physics, 100(2), 335–354. https://doi.org/10.1016/0021-9991(92)90240-y

7. Cerdeira, A. T. S., Campos, J. B. L. M., Miranda, J. M., & Araújo, J. D. P. (2020). Review on Microbubbles and Microdroplets Flowing through Microfluidic Geometrical Elements. Micromachines, 11(2), 201. https://doi.org/10.3390/mi11020201.

8. Chen, Y., & Deng, Z. (2017). Hydrodynamics of a droplet passing through a microfluidic T-junction. In Journal of Fluid Mechanics (Vol. 819, pp. 401–434). Cambridge University Press (CUP). https://doi.org/10.1017/jfm.2017.181

9. Chiu, D. T., deMello, A. J., Di Carlo, D., Doyle, P. S., Hansen, C., Maceiczyk, R. M., & Wootton, R. C. R. (2017). Small but Perfectly Formed? Successes, Challenges, and Opportunities for Microfluidics in the Chemical and Biological Sciences. Chem, 2(2), 201–223. https://doi.org/10.1016/j.chempr.2017.01.009.

10. Chong, Z. Z., Tan, S. H., Gañán-Calvo, A. M., Tor, S. B., Loh, N. H., & Nguyen, N.-T. (2016). Active droplet generation in microfluidics. Lab on a Chip, 16(1), 35–58. https://doi.org/10.1039/c5lc01012h.

11. Christopher, G. F., Noharuddin, N. N., Taylor, J. A., & Anna, S. L. (2008). Experimental observations of the squeezing-to-dripping transition in T-shaped microfluidic junctions. Physical Review E, 78(3). https://doi.org/10.1103/physreve.78.036317

12. Churski, K., Michalski, J., & Garstecki, P. (2010). Droplet on demand system utilizing a computer controlled microvalve integrated into a stiff polymeric microfluidic device. Lab Chip, 10(4), 512–518. https://doi.org/10.1039/b915155a.

13. Cui, P., & Wang, S. (2019). Application of microfluidic chip technology in pharmaceutical analysis: A review. Journal of Pharmaceutical Analysis, 9(4), 238–247. https://doi.org/10.1016/j.jpha.2018.12.001.

14. De Menech, M., Garstecki, P., Jousse, F., & Stone, H. A. (2008). Transition from squeezing to dripping in a microfluidic T-shaped junction. Journal of Fluid Mechanics, 595, 141–161. https://doi.org/10.1017/s002211200700910x.

15. Dolan, P. T., Whitfield, Z. J., & Andino, R. (2018). Mapping the Evolutionary Potential of RNA Viruses. Cell Host & Microbe, 23(4), 435–446. https://doi.org/10.1016/j.chom.2018.03.012.

16. Garstecki, P., Fuerstman, M. J., Stone, H. A., & Whitesides, G. M. (2006). Formation of droplets and bubbles in a microfluidic T-junction—scaling and mechanism of break-up. Lab on a Chip, 6(3), 437. https://doi.org/10.1039/b510841a

17. Glawdel, T., Elbuken, C., & Ren, C. L. (2012). Droplet formation in microfluidic T-junction generators operating in the transitional regime. I. Experimental observations. Physical Review E, 85(1). https://doi.org/10.1103/physreve.85.016322

18. Han, W., & Chen, X. (2019). Effect of Geometry Configuration on the Merged Droplet Formation in a Double T-Junction. In Microgravity Science and Technology (Vol. 31, Issue 6, pp. 855–864). Springer Science and Business Media LLC. https://doi.org/10.1007/s12217-019-09720-y

19. Huang, D., Wang, K., Wang, Y., Sun, H., Liang, X., & Meng, T. (2020). Precise control for the size of droplet in T-junction microfluidic based on iterative learning method. Journal of the Franklin Institute, 357(9), 5302–5316. https://doi.org/10.1016/j.jfranklin.2020.02.046.

20. Jamalabadi, M. Y. A., DaqiqShirazi, M., Kosar, A., & Shadloo, M. S. (2017). Effect of injection angle, density ratio, and viscosity on droplet formation in a microfluidic T-junction. Theoretical and Applied Mechanics Letters, 7(4), 243–251. https://doi.org/10.1016/j.taml.2017.06.002.

21. Kashid, M. N., Renken, A., & Kiwi-Minsker, L. (2010). CFD modelling of liquid–liquid multiphase microstructured reactor: Slug flow generation. Chemical Engineering Research and Design, 88(3), 362–368. https://doi.org/10.1016/j.cherd.2009.11.017

22. Korczyk, P. M., van Steijn, V., Blonski, S., Zaremba, D., Beattie, D. A., & Garstecki, P. (2019). Accounting for corner flow unifies the understanding of droplet formation in microfluidic channels. In Nature Communications (Vol. 10, Issue 1). Springer Science and Business Media LLC. https://doi.org/10.1038/s41467-019-10505-5

23. Köster, S., Angilè, F. E., Duan, H., Agresti, J. J., Wintner, A., Schmitz, C., Rowat, A. C., Merten, C. A., Pisignano, D., Griffiths, A. D., & Weitz, D. A. (2008). Drop-based microfluidic devices for encapsulation of single cells. Lab on a Chip, 8(7), 1110. https://doi.org/10.1039/b802941e.

24. Lan, F., Demaree, B., Ahmed, N., & Abate, A. R. (2017). Single-cell genome sequencing at ultra-high-throughput with microfluidic droplet barcoding. Nature Biotechnology, 35(7), 640–646. https://doi.org/10.1038/nbt.3880.

25. Li, X., He, L., He, Y., Gu, H., & Liu, M. (2019). Numerical study of droplet formation in the ordinary and modified T-junctions. Physics of Fluids, 31(8), 082101. https://doi.org/10.1063/1.5107425.

26. Li, X.-B., Li, F.-C., Yang, J.-C., Kinoshita, H., Oishi, M., & Oshima, M. (2012). Study on the mechanism of droplet formation in T-junction microchannel. Chemical Engineering Science, 69(1), 340–351. https://doi.org/10.1016/j.ces.2011.10.048.

27. Li, Y. K., Liu, G. T., Xu, J. H., Wang, K., & Luo, G. S. (2015). A microdevice for producing monodispersed droplets under a jetting flow. RSC Advances, 5(35), 27356–27364. https://doi.org/10.1039/c5ra02397a.

28. Li, Y. K., Wang, K., Xu, J. H., & Luo, G. S. (2016). A capillary-assembled micro-device for monodispersed small bubble and droplet generation. Chemical Engineering Journal, 293, 182–188. https://doi.org/10.1016/j.cej.2016.02.074.

29. Ling, Y., Fullana, J.-M., Popinet, S., & Josserand, C. (2016). Droplet migration in a Hele–Shaw cell: Effect of the lubrication film on the droplet dynamics. In Physics of Fluids (Vol. 28, Issue 6, p. 062001). AIP Publishing. https://doi.org/10.1063/1.4952398

30. Link, D. R., Anna, S. L., Weitz, D. A., & Stone, H. A. (2004). Geometrically Mediated Breakup of Drops in Microfluidic Devices. Physical Review Letters, 92(5). https://doi.org/10.1103/physrevlett.92.054503.

31. Liu, H., & Zhang, Y. (2009). Droplet formation in a T-shaped microfluidic junction. Journal of Applied Physics, 106(3), 034906. https://doi.org/10.1063/1.3187831

32. Mastiani, M., Mosavati, B., & Kim, M. (Mike). (2017). Numerical simulation of high inertial liquid-in-gas droplet in a T-junction microchannel. RSC Adv., 7(77), 48512–48525. https://doi.org/10.1039/c7ra09710g

33. Nekouei, M., & Vanapalli, S. A. (2017). Volume-of-fluid simulations in microfluidic T-junction devices: Influence of viscosity ratio on droplet size. Physics of Fluids, 29(3), 032007. https://doi.org/10.1063/1.4978801

34. Ngo, I.-L., Dang, T.-D., Byon, C., & Joo, S. W. (2015). A numerical study on the dynamics of droplet formation in a microfluidic double T-junction. Biomicrofluidics, 9(2), 024107. https://doi.org/10.1063/1.4916228.

35. Ngo, I.-L., Woo Joo, S., & Byon, C. (2016). Effects of Junction Angle and Viscosity Ratio on Droplet Formation in Microfluidic Cross-Junction. Journal of Fluids Engineering, 138(5). https://doi.org/10.1115/1.4031881.

36. Nisisako, T., & Torii, T. (2008). Microfluidic large-scale integration on a chip for mass production of monodisperse droplets and particles. Lab Chip, 8(2), 287–293. https://doi.org/10.1039/b713141k.

37. Raja, S., Satyanarayan, M. N., Umesh, G., & Hegde, G. (2021). Numerical Investigations on Alternate Droplet Formation in Microfluidic Devices. In Microgravity Science and Technology (Vol. 33, Issue 6). Springer Science and Business Media LLC. https://doi.org/10.1007/s12217-021-09917-0

38. Rivet, C., Lee, H., Hirsch, A., Hamilton, S., & Lu, H. (2011). Microfluidics for medical diagnostics and biosensors. Chemical Engineering Science, 66(7), 1490–1507. https://doi.org/10.1016/j.ces.2010.08.015.

39. Santos, R. M., & Kawaji, M. (2010). Numerical modeling and experimental investigation of gas–liquid slug formation in a microchannel T-junction. International Journal of Multiphase Flow, 36(4), 314–323. https://doi.org/10.1016/j.ijmultiphaseflow.2009.11.009

40. Sattari, A., Hanafizadeh, P., & Hoorfar, M. (2020). Multiphase flow in microfluidics: From droplets and bubbles to the encapsulated structures. Advances in Colloid and Interface Science, 282, 102208. https://doi.org/10.1016/j.cis.2020.102208.

41. Scheler, O., Postek, W., & Garstecki, P. (2019). Recent developments of microfluidics as a tool for biotechnology and microbiology. Current Opinion in Biotechnology, 55, 60–67. https://doi.org/10.1016/j.copbio.2018.08.004.

42. Shui, L., van den Berg, A., & Eijkel, J. C. T. (2009). Capillary instability, squeezing, and shearing in head-on microfluidic devices. Journal of Applied Physics, 106(12), 124305. https://doi.org/10.1063/1.3268364.

43. Singh, R., Bahga, S. S., & Gupta, A. (2020). Electrohydrodynamic droplet formation in a T-junction microfluidic device. Journal of Fluid Mechanics, 905. https://doi.org/10.1017/jfm.2020.749.

44. Soh, G. Y., Yeoh, G. H., & Timchenko, V. (2016). Numerical investigation on the velocity fields during droplet formation in a microfluidic T-junction. Chemical Engineering Science, 139, 99–108. https://doi.org/10.1016/j.ces.2015.09.025

45. Stone, H. A., Stroock, A. D., & Ajdari, A. (2004). ENGINEERING FLOWS IN SMALL DEVICES. Annual Review of Fluid Mechanics, 36(1), 381–411. https://doi.org/10.1146/annurev.fluid.36.050802.122124.

46. Sun, X., Zhu, C., Fu, T., Ma, Y., & Li, H. Z. (2018). Dynamics of droplet breakup and formation of satellite droplets in a microfluidic T-junction. Chemical Engineering Science, 188, 158–169. https://doi.org/10.1016/j.ces.2018.05.027.

47. Sussman, M., Smereka, P., & Osher, S. (1994). A Level Set Approach for Computing Solutions to Incompressible Two-Phase Flow. Journal of Computational Physics, 114(1), 146–159. https://doi.org/10.1006/jcph.1994.1155

48. Tan, S. H., & Nguyen, N.-T. (2011). Generation and manipulation of monodispersed ferrofluid emulsions: The effect of a uniform magnetic field in flow-focusing and T-junction configurations. Physical Review E, 84(3). https://doi.org/10.1103/physreve.84.03631.

49. Thorsen, T., Roberts, R. W., Arnold, F. H., & Quake, S. R. (2001). Dynamic Pattern Formation in a Vesicle-Generating Microfluidic Device. Physical Review Letters, 86(18), 4163–4166. https://doi.org/10.1103/physrevlett.86.4163.

50. van Sint Annaland, M., Deen, N. G., & Kuipers, J. A. M. (2005). Numerical simulation of gas bubbles behaviour using a three-dimensional volume of fluid method. Chemical Engineering Science, 60(11), 2999–3011. https://doi.org/10.1016/j.ces.2005.01.031

51. van Steijn, V., Kreutzer, M. T., & Kleijn, C. R. (2007). μ-PIV study of the formation of segmented flow in microfluidic T-junctions. Chemical Engineering Science, 62(24), 7505–7514. https://doi.org/10.1016/j.ces.2007.08.068

52. Viswanathan, H. (2019). Breakup and coalescence of drops during transition from dripping to jetting in a Newtonian fluid. International Journal of Multiphase Flow, 112, 269–285. https://doi.org/10.1016/j.ijmultiphaseflow.2018.09.016

53. Viswanathan, H. (2020). Oscillatory motion and merging responses of primary and satellite droplets from Newtonian liquid jets. Chemical Engineering Science, 212, 115334. https://doi.org/10.1016/j.ces.2019.115334

54. Wong, V.-L., Loizou, K., Lau, P.-L., Graham, R. S., & Hewakandamby, B. N. (2017). Numerical studies of shear-thinning droplet formation in a microfluidic T-junction using two-phase level-SET method. Chemical Engineering Science, 174, 157–173. https://doi.org/10.1016/j.ces.2017.08.027

55. Xu, J. H., Li, S. W., Tan, J., & Luo, G. S. (2008). Correlations of droplet formation in T-junction microfluidic devices: from squeezing to dripping. Microfluidics and Nanofluidics, 5(6), 711–717. https://doi.org/10.1007/s10404-008-0306-4

56. Xu, J. H., Li, S. W., Tan, J., Wang, Y. J., & Luo, G. S. (2006). Preparation of highly monodisperse droplet in a T-junction microfluidic device. AIChE Journal, 52(9), 3005–3010. https://doi.org/10.1002/aic.10924.

57. Youngs, D. (1982). Time-dependent multi-material flow with large fluid distortion. Numer. Methods Fluid Dyn. (1982), pp. 273-285.

58. Zhang, S., Liang, X., Huang, X., Wang, K., & Qiu, T. (2022). Precise and fast microdroplet size distribution measurement using deep learning. Chemical Engineering Science, 247, 116926. https://doi.org/10.1016/j.ces.2021.116926.

59. Zhao, C.-X., & Middelberg, A. P. J. (2011). Two-phase microfluidic flows. Chemical Engineering Science, 66(7), 1394–1411. https://doi.org/10.1016/j.ces.2010.08.038.

60. Zhu, P., & Wang, L. (2017). Passive and active droplet generation with microfluidics: a review. Lab on a Chip, 17(1), 34–75. https://doi.org/10.1039/c6lc01018k.

# Appendices

## Appendix A

To justify the choice of methods in capturing the interface of the two phases, a comparison between VOF and CLSVOF was undertaken as shown below in Fig.**14** with the fine mesh with the same hydrodynamic conditions and channel dimensions as in Fig.**2**.

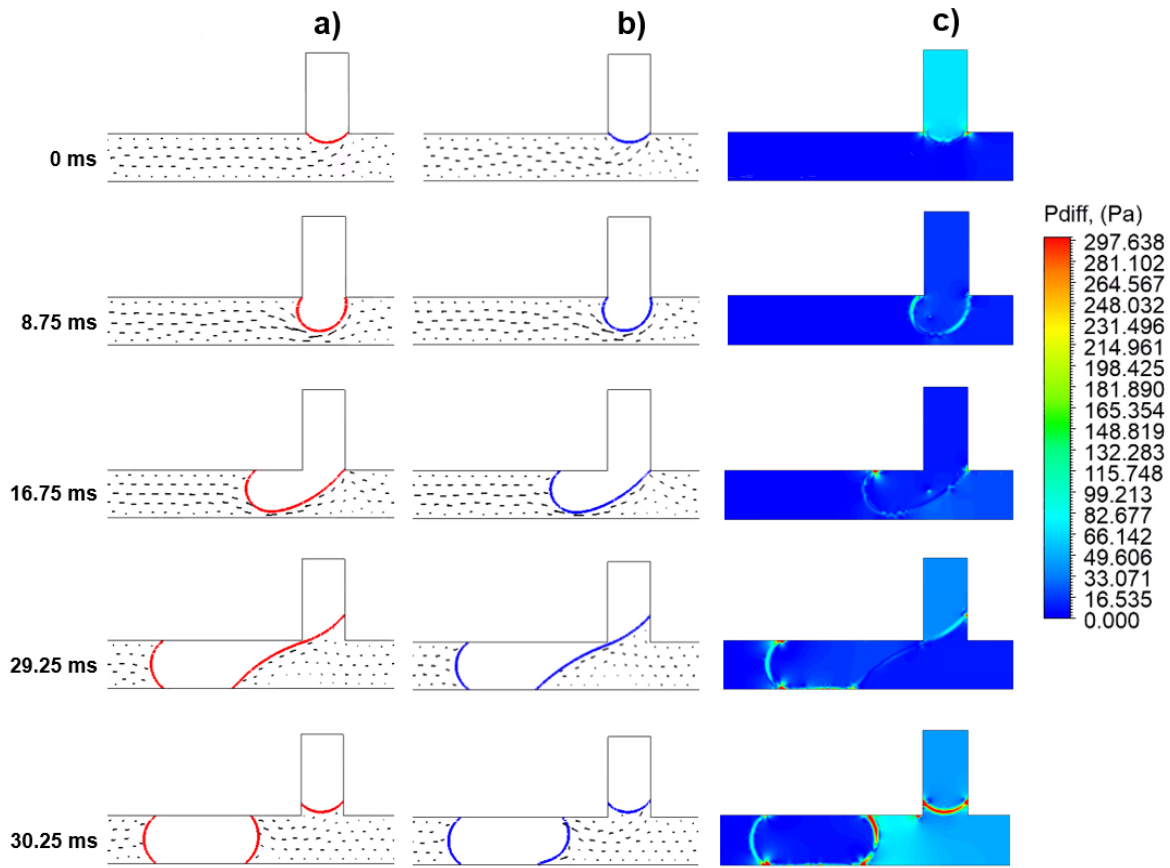

**Fig. 14**. A comparison between the volume fraction profiles obtained for $Ca$ = 0.0043, $q$ = 0.6487 using **a)** CLSVOF and **b)** VOF methods; both images are superimposed with the continuous phase velocity vectors. The figure **b)** shows the differences in the pressure distribution ($P_{diff}$= $P_{CLSVOF}$-$P_{VOF}$) in the channel between CLSVOF and VOF in the channel at the same instances.

The spatial and temporal evolution of the droplets' interface for both the CLSVOF and VOF methods appear to be identical. However, at the incipience of the breakup at 30.25 ms, subtle differences exist between these methods on the interface curvature, as seen in Fig.14**c**). The pressure difference $P_{diff}$, which is the difference in pressure predicted by the CLSVOF ($P_{CLSVOF}$) and the VOF ($P_{VOF}$), suggests that the CLSVOF predicts a marginally higher pressure at the interface just after the breakup. Nevertheless, such differences are negligible considering that the two important parameters for the current study, viz., a) the final drop shape and b) the frequency of drop formation predicted by the two methods, show differences of ~0.17% and ~0.02%, respectively suggesting excellent agreements between CLSVOF and VOF methods (**Table 2**).

**Appendix B**

An additional comparison is presented in **Fig.15** for different operating conditions to fortify the validation of the VOF model against the experimental data of Glawdel et al. (2012) using a fine grid size of 3 µm.

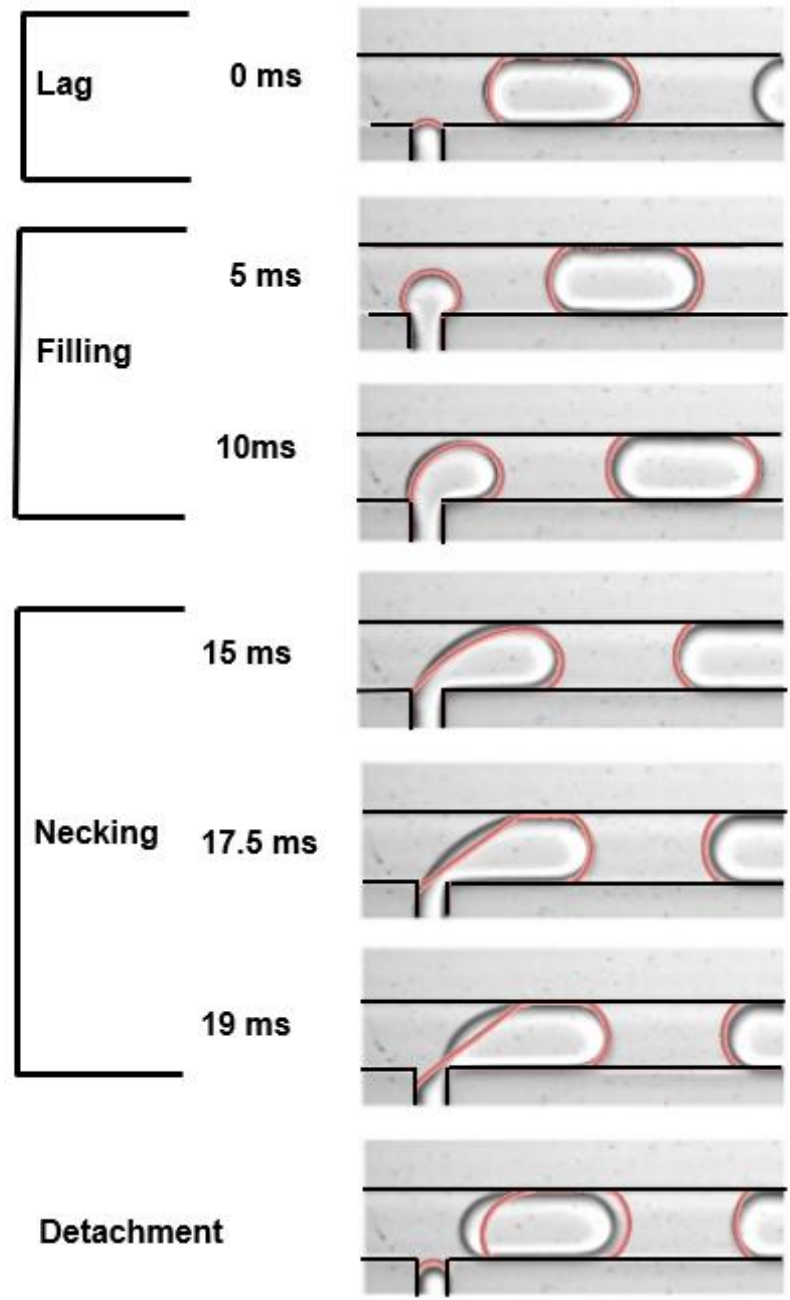

**Fig. 15**. Comparison between the experimental result of Glawdel et al. (2012) (Reproduced with permission from, Copyright 2012 APS) against the current numerical (VOF) predictions (shown by the solid red line) during different stages of drop formation for $Ca = 0.0027$, $q = 0.3517$.

In this case, $q = 0.3514$, parameters such as the width of the dispersed phase inlet ($W_d$) is 45 µm, and $Ca = 0.0027$ were maintained the same as the previously published experimental result of Glawdel et al. (2012). The remaining parameters were maintained the same as provided in **Table 1** of the manuscript. As shown in **Fig.15**, the drop formation frequency predicted by the numerical result is 45.454 Hz, whereas the experimental drop formation frequency reported was 45.8 Hz resulting in a difference of ~0.75%. The numerical results agree well with the experimental measurements for various stages of the drop formation process.